\documentclass[prd,aps,floats,twocolumn]{revtex4}

\usepackage{graphicx}
\usepackage{amssymb} 
\usepackage{amsmath}
\usepackage[T1]{fontenc} 
\usepackage[normalem]{ulem}
\usepackage{xcolor}

\begin{document}

\title{The Big Bang, CPT, and Neutrino Dark Matter}

\author{Latham Boyle$^1$, Kieran Finn$^{1,2}$ and Neil Turok$^1$}

\affiliation{$^1$Perimeter Institute for Theoretical Physics, \\ Waterloo, Ontario, Canada, N2L 2Y5 \\
$^2$School of Physics and Astronomy, University of Manchester \\ Manchester, UK, M13 9PL}  
   
\date{March 2019}

\begin{abstract}
  We investigate the idea that the universe before the Big Bang is the $CPT$ reflection of the universe after the bang, both classically and quantum mechanically, so that the universe does {\it not} spontaneously violate $CPT$.  We show how $CPT$ symmetry selects a preferred vacuum state for quantum fields on a $CPT$-invariant cosmological background spacetime. The universe before the bang and the universe after the bang may be viewed as a universe/anti-universe pair, emerging directly into the hot, radiation-dominated era we observe in our past.   This, in turn, leads to a remarkably economical explanation of the cosmological dark matter.  With no additional fields beyond Einstein gravity and the standard model of particle physics (including right-handed neutrinos), a $\mathbb{Z}_{2}$ symmetry stabilizes one of the right-handed neutrinos.  We calculate its abundance in detail and show that, in order to match the observed dark matter density, its mass must be $4.8\times10^{8}~{\rm GeV}$.  We obtain several further predictions, including: (i) that the three light neutrinos are majorana; (ii) that one of these is exactly massless; and (iii) that, in the absence of an epoch of cosmic inflation, there should be no primordial, long-wavelength gravitational waves. We also briefly discuss the natural origin of the matter-antimatter asymmetry within this picture and possibilities for explaining the cosmological perturbations. 
\end{abstract}                        

\maketitle 

\section{Introduction}

Cosmological observations indicate the universe to be astonishingly simple on the largest accessible scales \cite{Ade:2015xua, Aghanim:2015xee, Ade:2015hxq, Ade:2015ava, Ade:2015bva}. To an excellent approximation, we infer that, seconds after the Big Bang, the universe was accurately described by a spatially-flat, radiation-dominated FRW metric with small gaussian, adiabatic, scalar, growing perturbations with an almost scale-invariant power spectrum. So far, there is no evidence for primordial vector or tensor perturbations or cosmic defects.  The simplicity of the large-scale universe is commonly interpreted as evidence for a prior epoch of accelerated expansion called inflation \cite{Guth:1980zm, Linde:1981mu, Albrecht:1982wi, Kazanas:1980tx, Starobinsky:1980te, Sato:1981ds, Sato:1980yn}. However, inflationary models introduce a great deal of freedom which we would prefer to avoid. In this paper, and a companion letter \cite{Boyle:2018tzc}, we shall investigate a different possibility: that some aspects, at least, of the simple structure and content of the early universe may in fact be explained by $CPT$ symmetry, believed to be a fundamental symmetry of the laws of nature.  Among other things, we shall describe in detail a remarkable consequence of this hypothesis, namely a highly economical new explanation for the cosmological dark matter.

Our starting point is a simple observation. In the hot, radiation-dominated era the background metric is simply $g_{\mu\nu}\propto \tau^{2}\eta_{\mu\nu}$, where $\eta_{\mu\nu}$ is the Minkowski metric and $\tau$ is the conformal time. In addition to the usual cosmological (FRW) isometries, this metric has an additional isometry under $\tau\to-\tau$, {\it i.e.}, time reversal symmetry T.  We interpret this as a clue that the state of the universe ({\it i.e.}\ the spacetime itself as well as the quantum state of the QFT on that spacetime) might actually respect $CPT$ symmetry.  In this paper, we explore the general mathematical consequences of $CPT$ symmetry for both bosonic and fermionic fields living on an FRW background with time-reversal symmetry.  Our treatment significantly extends the standard treatment of $CPT$ symmetry from Minkowski spacetime~\cite{WeinbergQFT1} to FRW backgrounds.  In particular, we identify a $CPT$-invariant vacuum on FRW which generalizes the usual $CPT$-invariant Minkowski vacuum. Fascinatingly, for late time observers like us, this vacuum is not empty but has a finite density of particles. Thus, imposing $CPT$ symmetry on the universe results in a new ``production mechanism'' for cosmological dark matter, allowing the minimal standard model (including right handed neutrinos) to explain the dark matter without the need for any additional fields. 

In Minkowski spacetime, the natural vacuum is unique and unambiguous -- it is the one that respects the Minkowski isometries (more precisely: spacetime translations, Lorentz transformations, and $CPT$).  But in a more general curved spacetime, the choice of vacuum becomes ambiguous.  Different observers ({\it e.g.}\ inertial observers in different parts of the spacetime) will, in general, define different, inequivalent vacua so that the zero particle state according to one observer will contain particles according to a different observer~\cite{BirrellDavies, MukhanovWinitzki}.\footnote{This phenomenon is responsible for the Unruh radiation seen by an accelerated observer, the Hawking radiation produced by a black hole, and the primordial perturbations produced by inflation \cite{BirrellDavies, MukhanovWinitzki}.}  In particular, in an ordinary FRW spacetime, the isometries (spatial translations, spatial rotations, and parity) are not enough to determine a preferred vacuum, and observers at different epochs will disagree.  But, as we explain below, once we follow our own cosmological background through the Bang, endowing it with an extra isometry (time reversal), we are led to a preferred vacuum that respects the full isometry group of the background (more precisely: spatial translations, spatial rotations and $CPT$).  In other words: it becomes possible to adopt the natural hypothesis that the state of our universe does {\it not} spontaneously violate $CPT$: this simple hypothesis is satisfied by the background, and also selects a state for the quantum fields on that background.

Assuming spatial homogeneity for the moment, let $n_{j}(t,{\bf p},h)$ denote the momentum space distribution function of particles of species $j$ and helicity $h$; and let $n_{j}^{c}(t,{\bf p},h)$ denote the distribution function for the corresponding anti-particles.  Under CPT, we have:
\begin{equation}
  n_{j}(t,{\bf p},h)
  \!\xrightarrow{\!C\!}\!n_{j}^{c}(t,{\bf p},h)
  \!\xrightarrow{\!P\!}\!n_{j}^{c}(t,-{\bf p},\!-h)
  \!\xrightarrow{\!T\!}\!n_{j}^{c}(-t,{\bf p},\!-h).
\end{equation}
So CPT invariance implies that the cosmological distribution functions satisfy
\begin{equation}
  \label{CPT_constraint}
  n_{j}(t,{\bf p},h)=n_{j}^{c}(-t,{\bf p},-h).
\end{equation}
In other words, the density of particles of species $j$ with momentum ${\bf p}$ and helicity $h$ at time $t$ {\it after} the bang equals the density of the corresponding anti-particle species with momentum ${\bf p}$ and helicity $-h$ at time $-t$ {\it before} the bang.  Thus, if the universe after the bang has a slight excess of matter, the universe before the bang has a slight excess of anti-matter (see \cite{Albrow:1973wp} for an early thought in this direction).  Moreover, as explained in a companion paper \cite{Boyle:2018tzc}, density perturbations grow as we get further from the bang in either direction, and hence the physical (thermodynamic) arrow of time points {\it away} from the bang in both directions (to the future and past).  Recalling the Stueckelberg interpretation of an anti-particle as a particle running backward in time \cite{Stueckelberg:1941, Stueckelberg:1941rg, Feynman:1949hz}, we are naturally led to reinterpret our $CPT$-symmetric universe as a universe/anti-universe pair, emerging from nothing \cite{Boyle:2018tzc}!  

Now, if we assume that the matter fields in the universe are described by the standard model of particle physics (including a right-handed neutrino in each generation), then there is only one possible dark matter candidate -- namely, one of the right-handed (sterile) neutrinos.  Ordinarily, in the same limit that this particle becomes stable, it also becomes decoupled from all of the other particles in the standard model, and hence is not produced by the thermal bath in the early universe.  But in our picture, it is still produced gravitationally: the mere fact that the universe is in the $CPT$-invariant state implies that this neutrino has a non-zero abundance (according to late-time observers like us).  We show that this particle accounts for the observed dark matter if its mass is $4.8\times10^{8}~{\rm GeV}$.  The other two heavy neutrinos are unstable and decay in the early universe. Their decays can naturally produce the observed matter/anti-matter asymmetry via the usual leptogenesis mechanism \cite{Fukugita:1986hr, Buchmuller:2005eh}.  This line of thought also makes other observable predictions.  Here we mention three: (i) that the three light (active) neutrinos are Majorana; (ii) that the lightest neutrino is massless; and (iii) that (in the absence of an inflationary epoch prior to the radiation era) no long-wavelength primordial gravitational waves are generated.

We think this scenario provides an appealing and very economical new picture which explains a number of key features of our universe.   Rather than adding fields, we impose symmetries: $CPT$, local scale invariance, and a discrete $\mathbb{Z}_{2}$ (to stabilize the dark matter neutrino). 

In this paper, we analytically extend the metric across the bang.  In a recent follow-up paper \cite{Boyle:2021jej}, we {\it also} analytically extend it  across the (de-Sitter-like) future  boundary of the spacetime; we show that, if we take the symmetries and complex analytic structure of this maximally extended spacetime seriously, and demand that the fields living on the spacetime respect this structure, then we are forced to impose a reflecting boundary condition at the bang, but no such boundary condition at future infinity. This, in turn, explains several observed properties of the primordial perturbations, provides a simple new and fundamental explanation for the observed arrow of time, {\it i.e.}, the fact that entropy increases as we get further from the bang, and suggests a new proposal for the wavefunction of the universe.  Here we add that the two-sheeted picture of spacetime obtained in \cite{Boyle:2021jej} provides an important theoretical underpinning for the CPT-symmetric vacuum state advocated in the present paper: if the matter fields respect the symmetry of the extended spacetime under swapping the two sheets, their vacuum should too.

At this stage of our understanding of a $CPT$-symmetric universe, we do not claim to fully explain the homogeneity, isotropy and flatness of the universe, nor the power spectrum of the primordial density perturbations, although steps toward all of these questions are taken in \cite{Boyle:2018tzc, Boyle:2021jej, Boyle:2021jaz}. 

The outline of this paper is as follows.  In Section 2, as a warm-up, we construct the CPT invariant vacuum state for a complex scalar field on a time-reversal-symmetric FRW background.  In Section 3, we then construct the CPT invariant vacuum state for a spin 1/2 fermion field.  A surprising new wrinkle appears here (which, as far as we are aware, has not been previously noticed): the fermion's effective mass $\mu(\tau)$ (in Minkowski gauge) may be either an even or odd function of $\tau$.  This leads to two different flavors of time-reversal symmetry in the fermion sector: the one that is found in QFT textbooks, and the one that is actually relevant to cosmology. In Sections 4 and 5, we apply these ideas to the standard model of particle physics and point out how they yield a new explanation for the observed dark matter.  
More specifically, we show: (i) that one of the three right-handed neutrinos becomes a good dark matter candidate; (ii) that if its mass is $4.8\times10^{8}~{\rm GeV}$, its abundance matches the presently observed dark matter abundance; (iii) that this dark matter candidate is automatically {\it ultra-cold}, with perturbations that are automatically {\it adiabatic} (in agreement with observations); (iv) that this perspective also gives a new view of the cosmic matter/anti-matter asymmetry; and (v) that this line of reasoning also makes three other predictions (about the neutrino sector and about primordial gravitational waves) which will be tested by forthcoming experiments.  In Section 6, we summarize our results, and discuss interesting connections to other theoretical ideas and experimental tests as well as topics for future work.

\section{$CPT$ invariant vacuum in FRW: scalar field}  

In this section we warm up with a scalar field.  Our treatment will be slightly pedantic at points to prepare for the analogous but somewhat subtler spinor story in the next section.

\subsection{The in/out bases}
\label{mode_expansion_scalar}

Consider a flat FRW background:
\begin{equation}
  ds^{2}=a^{2}(\tau)[-d\tau^{2}+d{\bf x}^{2}]
\end{equation}
and a complex scalar field $\Phi$ with mass $m>0$ on this background.  (The case of a real scalar field can be handled by a straightforward specialization of the following analysis.)  Note: throughout this paper, we will use coordinates $x=(\tau,{\bf x})$ where $\tau$ is the conformal time and ${\bf x}$ is the comoving spatial coordinate.   The Lagrangian is
\begin{subequations}
  \begin{eqnarray}
    \label{L_phi}
    L&=&\sqrt{-g}\big[-g^{\mu\nu}(\partial_{\mu}\Phi)^{\dagger}(\partial_{\nu}\Phi)
    -m^{2}\Phi^{\dagger}\Phi\big] \\
    &=&(\varphi')^{\dagger}(\varphi')-(\nabla\varphi)^{\dagger}(\nabla\varphi)-\mu^{2}\varphi^{\dagger}\varphi
  \end{eqnarray}
\end{subequations}
where, for convenience, in the second line we have introduced the Weyl invariant scalar field $\varphi=\varphi({\bf x},\tau)$ and its effective mass squared $\mu^{2}=\mu^{2}(\tau)$:
\begin{equation}
  \label{mu_scalar}
  \varphi\equiv a\Phi\qquad{\rm and}\qquad\mu^{2}\equiv(am)^{2}-a''/a.
\end{equation}
From here we obtain the equation of motion
\begin{equation}
  \label{KG_eq}
  \varphi''-\nabla^{2}\varphi+\mu^{2}\varphi=0,
\end{equation}
where a prime denotes $d/d\tau$.  Because of the FRW symmetry of the background, if $\varphi({\bf x},\tau)$ is a solution of (\ref{KG_eq}), so is:
\begin{itemize}
\item its spatial translation 
\begin{subequations}
  \begin{equation}
    \label{phi_y}
    \varphi_{{\bf y}}({\bf x},\tau)\equiv\varphi({\bf x}+{\bf y},\tau), 
  \end{equation}
  \item its spatial rotation 
  \begin{equation}
    \varphi_{R}({\bf x},\tau)\equiv\varphi(R{\bf x},\tau),\;\;\;\;
  \end{equation}
  \item its charge conjugate 
  \begin{equation}
    \label{phi_c}
    \varphi_{c}({\bf x},\tau)\equiv\varphi^{\ast}({\bf x},\tau),\;\;\;\!\;\;
  \end{equation}
  \item its parity reverse
  \begin{equation} 
    \label{phi_p}
    \varphi_{p}({\bf x},\tau)\equiv\varphi(-{\bf x},\tau),\;\;\!\;\;
  \end{equation}
  \item and (if $a^{2}$ and hence $\mu^{2}$ is an even function of $\tau$) its time reverse
  \begin{equation}
    \label{phi_t}
    \varphi_{t}({\bf x},\tau)\equiv\varphi^{\ast}({\bf x},-\tau).\;\;
  \end{equation}
\end{subequations}
\end{itemize}

We would like to expand the field $\varphi(x)$ in a basis of solutions of Eq.~(\ref{KG_eq}).  Let us consider how the various symmetries of the FRW background effect this expansion:
\begin{itemize}
\item {\it Spatial translations.}  Because of the spatial translation invariance, we can take the solutions to be spatial fourier modes: $\varphi({\bf x},\tau)=\varphi({\bf p},\tau){\rm e}^{i{\bf p}{\bf x}}$.  Passing to fourier space, the equation of motion (\ref{KG_eq}) becomes
\begin{equation}
  \label{KG_eq_fourier}
  \varphi''+(p^{2}+\mu^{2})\varphi=0\qquad(p\equiv|{\bf p}|).
\end{equation}
\item {\it Charge conjugation.}  Since (\ref{KG_eq_fourier}) is a second order equation, there are two linearly independent solutions per wavenumber ${\bf p}$.  We can take these two solutions to be the (positive frequency) solution $\varphi({\bf p},x)=u({\bf p},\tau){\rm e}^{i{\bf p}{\bf x}}$ and its charge conjugate (negative frequency) solution $\varphi_{c}({\bf p},x)=u^{\ast}({\bf p},\tau){\rm e}^{-i{\bf p}{\bf x}}$, and expand the field as follows
\begin{subequations}
  \begin{eqnarray}
    \label{phi_expansion_1}
    \varphi(x)\!&\!=\!&\!\int\frac{d^{3}p}{(2\pi)^{3}}\left[a({\bf p})\varphi({\bf p},x)\right. \nonumber\\
    \!&\!\!&\!\qquad\qquad\;\left.+b^{\dagger\!}({\bf p})\varphi_{c}({\bf p},x)\right] \\
    \label{phi_expansion_2}
    \!&\!=\!&\!\int\frac{d^{3}p}{(2\pi)^{3}}\left[a({\bf p})u({\bf p},\tau){\rm e}^{+i{\bf p}{\bf x}}\right. \nonumber\\
    \!&\!\!&\!\qquad\qquad\;\left.+b^{\dagger\!}({\bf p})u^{\ast}\!({\bf p},\tau){\rm e}^{-i{\bf p}{\bf x}}\right].
  \end{eqnarray}
\end{subequations}
Note that we have written $b^{\dagger}$ instead of $b^{\ast}$, in anticipation of quantization below.
\item \sout{{\it Time translations.}} Because FRW does {\it not} have time-translation symmetry, the ingoing (early time) and outgoing (late time) observers will generally disagree about which solutions have "positive frequency", and so they will perform the mode expansion (\ref{phi_expansion_2}) in two different ways:
\begin{eqnarray}
  \label{phi_expansion_pm}
  \varphi(x)\!&\!=\!&\!\int\frac{d^{3}p}{(2\pi)^{3}}\left[a_{\pm}^{}({\bf p})u_{\pm}^{}({\bf p},\tau){\rm e}^{+i{\bf p}{\bf x}} \right. \nonumber\\
  \!&\!\!&\!\qquad\qquad\;\left.+b_{\pm}^{\dagger}({\bf p})u_{\pm}^{\ast}({\bf p},\tau){\rm e}^{-i{\bf p}{\bf x}}\right].
\end{eqnarray}
Here the "$-$" and "$+$" subscripts correspond to the ingoing and outgoing observers, respectively.  The ingoing and outgoing bases are fixed by taking the positive frequency modes $u_{\pm}({\bf p},\tau)$ to satisfy the boundary conditions
\begin{equation}
  \label{scalar_bc}
  u_{\pm}({\bf p},\tau)\!\to\!\frac{1}{\sqrt{2\omega(p,\tau)}}{\rm exp}\left[-i\int_{\pm\tau_{0}^{}(p)}^{\tau}\!\!\!\!\!\!\!\!\!\!\!\!\omega(p,\tilde{\tau})d\tilde{\tau}\right]\!\!\!\!\!\!
\end{equation}
as $\tau\to\pm\infty$, where $\omega(p,\tau)>0$ is the positive root of $\omega^{2}\equiv p^{2}+\mu^{2}$, while $\tau_{0}^{}(p)$ is arbitrary, and may be fixed for convenience (we return to this point in Subsection \ref{scalar_bogoliubov_section}).  
\item {\it Spatial rotations.}  Since we have chosen a boundary condition (\ref{scalar_bc}) that respects the spatial rotational invariance of the equation of motion (\ref{KG_eq_fourier}), it follows that the solution $u_{\pm}({\bf p},\tau)$ is independent of the direction of ${\bf p}$
\begin{equation}  
  \label{scalar_isotropy}
  u_{\pm}({\bf p},\tau)=u_{\pm}(p,\tau).
\end{equation}

\item {\it Parity.}  If $\varphi(x)=u_{\pm}({\bf p},\tau){\rm e}^{i{\bf p}{\bf x}}$ is an outgoing (ingoing) positive frequency solution with momentum ${\bf p}$, then $\varphi_{p}(x)$ is another outgoing (ingoing) positive frequency solution with momentum $-{\bf p}$: $\varphi_{p}(x)\propto u_{\pm}(-{\bf p},\tau){\rm e}^{i(-{\bf p}){\bf x}}$.  Thus $u_{\pm}({\bf p},\tau)\propto u_{\pm}(-{\bf p},\tau)$ and, in particular, (\ref{scalar_bc}) implies
\begin{equation} 
  \label{upm_P}
  u_{\pm}({\bf p},\tau)=u_{\pm}(-{\bf p},\tau).
\end{equation}
In the scalar case, this parity constraint is redundant: it is already implied by (\ref{scalar_isotropy}).

\item {\it Time reversal.}  If $a^{2}(\tau)$ is an even function of $\tau$ and $\varphi(x)=u_{\pm}({\bf p},\tau){\rm e}^{i{\bf p}{\bf x}}$ is an outgoing (ingoing) positive frequency solution with momentum ${\bf p}$, then $\varphi_{t}(x)$ is an ingoing (outgoing) positive frequency solution with momentum $-{\bf p}$: $\varphi_{t}(x)\propto u_{\mp}^{}(-{\bf p},\tau){\rm e}^{i(-{\bf p}){\bf x}}$.  Thus $u_{\pm}^{\ast}({\bf p},-\tau)\propto u_{\mp}^{}(-{\bf p},\tau)$ and, in particular, (\ref{scalar_bc}) implies
\begin{equation}
  \label{upm_T}
  u_{\pm}^{\ast}({\bf p},-\tau)=u_{\mp}^{}(-{\bf p},\tau).
\end{equation}
\end{itemize}

\subsection{Canonical quantization in the in/out bases}
\label{canon_quant_scalar}

To quantize, we take the field $\varphi$ (\ref{phi_expansion_pm}) and its conjugate momentum
\begin{equation}
  \pi=\frac{\partial L}{\partial\varphi'}=(\varphi')^{\dagger}
\end{equation}
to satisfy the canonical commutation relations 
\begin{equation}
  \label{canon_comm_scalar}
    [\varphi({\bf x},\tau),\pi({\bf y},\tau)]=i\delta({\bf x}-{\bf y})\quad({\rm all}\;{\rm others}\;{\rm vanish}).
\end{equation}
If we note that the mode functions $u_{\pm}({\bf p},\tau)$ satisfy the normalization condition
\begin{equation}
  \label{Wronskian_pm}
  u_{\pm}^{}({\bf p},\tau)u_{\pm}^{\ast}{}'({\bf p},\tau)-u_{\pm}^{\ast}(-{\bf p},\tau)u_{\pm}^{}{}'(-{\bf p},\tau)=i,
\end{equation}
we can check that the canonical commutation relations (\ref{canon_comm_scalar}) are equivalent to the standard commutation relations for creation and annihilation operators:
\begin{equation}
  \label{ab_comm_pm}
  [a_{\pm}^{}({\bf p}),a_{\pm}^{\dagger}({\bf q})]=
  [b_{\pm}^{}({\bf p}),b_{\pm}^{\dagger}({\bf q})]=\delta({\bf p}-{\bf q})
\end{equation}
(all others vanish). The interpretation is that $a^{\dagger}({\bf p})$ creates a particle with momentum ${\bf p}$ while $b^{\dagger}({\bf p})$ creates the corresponding anti-particle with momentum ${\bf p}$.

Now let us see how these operators transform under $C$, $P$ and $T$ (see Ref.~\cite{WeinbergQFT1}):
\begin{itemize}
\item $C$:  The requirement that $\varphi$ should transform under charge conjugation like
\begin{equation}
  C\varphi(x)C^{-1}=\xi_{c}^{\ast}\varphi_{c}(x)
\end{equation}
(where $C$ is unitary and $\xi_{c}$ is the associated charge conjugation phase) implies that the creation and annihilation operators 
in (\ref{phi_expansion_pm}) transform like
\begin{subequations}  
  \label{C_scalar}
  \begin{eqnarray}  
    Ca_{\pm}({\bf p})C^{-1}&=&\xi_{c}^{\ast}b_{\pm}({\bf p}), \\
    Cb_{\pm}^{\dagger}({\bf p})C^{-1}&=&\xi_{c}^{\ast}a_{\pm}^{\dagger}({\bf p}).
  \end{eqnarray}
\end{subequations}

\item $P$: Eq.~(\ref{upm_P}) along with the requirement that $\varphi$ should transform under parity like
\begin{equation}
  P\varphi(x)P^{-1}=\xi_{p}^{\ast}\varphi_{p}(x),
\end{equation}
(where $P$ is unitary and $\xi_{p}$ is the associated parity phase) implies that the creation and annihilation operators 
in (\ref{phi_expansion_pm}) transform like
\begin{subequations} 
  \label{P_scalar} 
  \begin{eqnarray}
    Pa_{\pm}({\bf p})P^{-1}&=&\xi_{p}^{\ast}a_{\pm}(-{\bf p}), \\
    Pb_{\pm}^{\dagger}({\bf p})P^{-1}&=&\xi_{p}^{\ast}b_{\pm}^{\dagger}(-{\bf p}).
  \end{eqnarray}
\end{subequations}

\item $T$:  Eq.~(\ref{upm_T}) plus the requirement that $\varphi$ should transform under time-reversal like
\begin{equation}
  T\varphi(x)T^{-1}=\xi_{t}^{\ast}\varphi_{t}^{\ast}(x),
\end{equation}
(where $T$ is anti-unitary and $\xi_{t}$ is the associated time-reversal phase) implies that the creation and annihilation 
operators in (\ref{phi_expansion_pm}) transform like 
\begin{subequations}
  \label{T_scalar}
  \begin{eqnarray}
    Ta_{\pm}({\bf p})T^{-1}&=&\xi_{t}^{\ast}a_{\mp}(-{\bf p}), \\
    Tb_{\pm}^{\dagger}({\bf p})T^{-1}&=&\xi_{t}^{\ast}b_{\mp}^{\dagger}(-{\bf p}).
  \end{eqnarray}
\end{subequations}
\end{itemize}

\subsection{Transformation between the in/out bases}
\label{scalar_bogoliubov_section}

The spatial translation invariance of (\ref{KG_eq}) implies that an ingoing solution with spatial dependence ${\rm e}^{i{\bf p}{\bf x}}$ must evolve into an outgoing solution with spatial dependence ${\rm e}^{i{\bf p}{\bf x}}$.  Thus, the ingoing positive frequency solution $u_{-}({\bf p},\tau){\rm e}^{i{\bf p}{\bf x}}$ must be a linear combination of the outgoing positive and negative frequency solutions $u_{+}({\bf p},\tau){\rm e}^{i{\bf p}{\bf x}}$ and $u_{+}^{\ast}(-{\bf p},\tau){\rm e}^{i{\bf p}{\bf x}}$.  

Recalling, from Eq.~(\ref{scalar_isotropy}), that $u_{\pm}({\bf p},\tau)=u_{\pm}(p,\tau)$, we can write this linear combination as follows:
\begin{equation}
  \label{um_from_up}
  u_{-}(p,\tau)=\alpha(p)u_{+}(p,\tau)+\beta(p)u_{+}^{\ast}(p,\tau).
\end{equation}
The time-reversal condition (\ref{upm_T}) then implies that $|\alpha(p)|^{2}-|\beta(p)|^{2}=1$ and that $\beta(p)$ is imaginary.  Then, adjusting the choice of $\tau_{0}^{}(p)$ in (\ref{scalar_bc}) adjusts $u_{\pm}(p,\tau)\to{\rm e}^{\pm i\chi(p)}u_{\pm}(p,\tau)$, and we use this freedom to phase-rotate $\alpha(p)$ to make it real and non-negative.  Putting this all together, we can write 
\begin{equation}
  \alpha(p)={\rm cosh}[\lambda(p)]\qquad{\rm and}\qquad\beta(p)=i\,{\rm sinh}[\lambda(p)],
\end{equation}
where $\lambda(p)$ is real.  Now if we substitute (\ref{um_from_up}) into (\ref{phi_expansion_pm}), we see that the Bogoliubov transformation 
\begin{equation}
  \label{scalar_bogoliubov}
  \left[\begin{array}{c} a_{+}^{}(+{\bf p}) \\ b_{\;\!+}^{\dagger}(-{\bf p}) \end{array}\right]=B(p)
   \left[\begin{array}{c} a_{-}^{}(+{\bf p}) \\ b_{\;\!-}^{\dagger}(-{\bf p})\end{array}\right]
\end{equation}
between the "in" operators ($a_{-}$ and $b_{-}$) and the "out" operators ($a_{+}$ and $b_{+}$) is encoded in the 
Bogoliubov matrix
\begin{equation}
  \label{B_scalar}
    B(p)=\left[\begin{array}{rr} {\rm cosh}\,\lambda(p)\quad& -i\,{\rm sinh}\,\lambda(p) \\
   +i\,{\rm sinh}\,\lambda(p)\quad&{\rm cosh}\,\lambda(p)\end{array}\right].
\end{equation}

The Bogoliubov transformation (\ref{scalar_bogoliubov}) makes sense physically.  Since the (FRW) background breaks neither internal $U(1)$ invariance, nor spatial translational invariance, the Bogoliubov transformation should "conserve" charge and spatial momentum: {\it i.e.}\ the "out" operator $a_{+}({\bf p})$ (which annihilates a particle of momentum ${\bf p}$) can only be a linear combination of the "in" operator $a_{-}({\bf p})$ (which also annihilates a particle of momentum ${\bf p}$) and the "in" operator 
$b_{-}^{\dagger}(-{\bf p})$ (which creates an anti-particle of momentum $-{\bf p}$).

The function $\lambda(p)$ measures the physical "offset" between the operators $\{a_{-},b_{-}\}$ that annihilate the "in" vacuum $|0_{-}\rangle$ and the operators $\{a_{+},b_{+}\}$ that annihilate the "out" vacuum $|0_{+}\rangle$:
\begin{subequations}
  \begin{eqnarray}
    a_{-}({\bf p})|0_{-}\rangle=b_{-}({\bf p})|0_{-}\rangle&=&0, \\
    a_{+}({\bf p})|0_{+}\rangle=b_{+}({\bf p})|0_{+}\rangle&=&0.
  \end{eqnarray}
\end{subequations}
Unless $\lambda(p)$ vanishes for all $p$, the "in" and "out" vacua are inequivalent: {\it e.g.}\ the "out" observer's number operator $a_{+}^{\dagger}({\bf p})a_{+}({\bf p})$ will have non-zero expectation value in the "in" observer's vacuum $|0_{-}\rangle$, and vice versa.  In this case, even if the {\it background} is invariant under $CPT$, the corresponding "in" and "out" vacua are not:
\begin{subequations}
  \begin{eqnarray}
    CPT|0_{-}\rangle&\propto&|0_{+}\rangle, \\
    CPT|0_{+}\rangle&\propto&|0_{-}\rangle.  
  \end{eqnarray}
\end{subequations}

\subsection{$CPT$ invariant bases and vacua}
\label{CPT_vacua_scalar}

We have seen that the "in" and "out" vacua $|0_{+}\rangle$ and $|0_{-}\rangle$ are {\it not} $CPT$ invariant.  In this subsection we construct all the vacua that {\it are} $CPT$ invariant.

Ultimately, it was the requirement (\ref{scalar_bc}) that the basis modes $u$ have purely positive frequency (in either the far past or the far future) that forced us to introduce two inequivalent bases ("in" and "out") -- and hence two inequivalent vacua ($|0_{-}\rangle$ and $|0_{+}\rangle$) -- that are swapped by $CPT$.  To construct bases and vacua that are {\it preserved} by $CPT$, we must give up this requirement.  We replace the two expansions (\ref{phi_expansion_pm}) by the single expansion (\ref{phi_expansion_2}), while the conditions (\ref{scalar_isotropy}), (\ref{upm_P}) and (\ref{upm_T}) are replaced by
\begin{eqnarray}
  \label{u0_scalar_isotropy}
  u({\bf p},\tau)&=&u(p,\tau), \\ 
  \label{u0_P}
  u({\bf p},\tau)&=&u(-{\bf p},\tau), \\
  \label{u0_T}
  u({\bf p},\tau)&=&u^{\ast}(-{\bf p},-\tau).
\end{eqnarray}

Once again, the canonical equal-time commutation relations (\ref{canon_comm_scalar}) are equivalent to the 
usual commutation relations for creation and annihilation operators
\begin{equation}
  \label{ab_comm_0}
  [a({\bf p}),a^{\dagger}({\bf q})]=[b({\bf p}),b^{\dagger}({\bf q})]=(2\pi)^{3}\delta({\bf p}-{\bf q})
\end{equation}
provided the mode functions satisfy the Wronskian normalization condition 
\begin{equation}
  \label{Wronskian_0}
  u({\bf p},\tau)u^{\ast}{}'({\bf p},\tau)-u^{\ast}(-{\bf p},\tau)u'(-{\bf p},\tau)=i.
\end{equation}

Now, we can re-express the solution $u({\bf p},\tau){\rm e}^{i{\bf p}{\bf x}}$ as a linear combination of either the ingoing solutions,
$u_{-}^{}({\bf p},\tau){\rm e}^{i{\bf p}{\bf x}}$ and $u_{-}^{\ast}(-{\bf p},\tau){\bf e}^{i{\bf p}{\bf x}}$, or the outgoing solutions,
$u_{+}^{}({\bf p},\tau){\rm e}^{i{\bf p}{\bf x}}$ and $u_{+}^{\ast}(-{\bf p},\tau){\bf e}^{i{\bf p}{\bf x}}$.  Since $u({\bf p},\tau)=u(p,\tau)$
and $u_{\pm}({\bf p},\tau)=u_{\pm}(p,\tau)$, we can write these linear combinations as follows
\begin{equation}
  \label{u_from_upm}
  u(p,\tau)=\alpha_{\pm}^{}(p)u_{\pm}^{}(p,\tau)+\beta_{\pm}^{}(p)u_{\pm}^{\ast}(p,\tau).
\end{equation}
If we substitute (\ref{u_from_upm}) into (\ref{phi_expansion_2}) and compare to (\ref{phi_expansion_pm}), we see that the Bogoliubov transformations
\begin{equation}
  \label{Bogoliubov_pm_scalar}
  \left[\begin{array}{c} a_{\pm}^{}(+{\bf p}) \\ b_{\pm}^{\dagger}(-{\bf p}) \end{array}\right]=
  B_{\pm}(p)
  \left[\begin{array}{r} a\,(+{\bf p}) \\ b_{}^{\dagger\!}(-{\bf p}) \end{array}\right]
\end{equation}
from the operators $\{a,b\}$, to the "in" or "out" operators, $\{a_{-},b_{-}\}$ or $\{a_{+},b_{+}\}$,  are described, respectively, by the Bogliubov matrices $B_{-}$ and $B_{+}$:
\begin{equation}
  \label{B_pm_scalar}
  B_{\pm}(p)\equiv\left[\begin{array}{cc} \alpha_{\pm}^{}(p) & \quad\beta_{\pm}^{\ast}(p) \\
  \beta_{\pm}^{}(p) & \quad\alpha_{\pm}^{\ast}(p)\end{array}\right].
\end{equation}
Note that $B_{+}$ and $B_{-}$ must obey the following three constraints:
\begin{enumerate}
\begin{subequations}
\item First, compatibility between the commutation relations (\ref{ab_comm_pm}) and (\ref{ab_comm_0}) imply
\begin{equation}
  \label{alpha_beta_pm_normalization}
  {\rm Det}[B_{\pm}(p)]=1.
\end{equation}
\item Second, compatibility between the time-reversal constraints (\ref{upm_T}) and (\ref{u0_T}) imply
\begin{equation}
  \label{alpha_beta_pm_T}
  B_{-}(p)=B_{+}^{\ast}(p).
\end{equation}
\item Third, the map $B(p)$ in Eq.~(\ref{B_scalar}), from the "in" operators $\{a_{-},b_{-}\}$ to the "out" operators
$\{a_{+},b_{+}\}$, can be re-expressed in terms of $B_{+}$ and $B_{-}$ as follows
\begin{equation}
  B(p) = B_{+}(p)B_{-}^{-1}(p).
\end{equation}
\end{subequations}
\end{enumerate}
The most general solution for $B_{\pm}$ satisfying these three constraints is
\begin{equation}
   \label{B_pm_explicit_scalar}
  B_{\pm}(p)=\hat{B}_{\pm}(p)B_{0}(p)
\end{equation}
where
\begin{subequations}  
  \begin{eqnarray}
    \hat{B}_{\pm}(p)&\equiv&\left[\begin{array}{rr}
    {\rm cosh}\frac{\lambda(p)}{2} & \quad \mp i\,{\rm sinh}\frac{\lambda(p)}{2} \\
    \pm i\,{\rm sinh}\frac{\lambda(p)}{2} & {\rm cosh}\frac{\lambda(p)}{2} \end{array}\right], \\
    B_{0}(p)&\equiv&\left[\begin{array}{rr}
    {\rm cosh}\,\eta(p) & \quad{\rm sinh}\,\eta(p) \\
    {\rm sinh}\,\eta(p) & \quad{\rm cosh}\,\eta(p) \end{array}\right].
  \end{eqnarray}
\end{subequations}
Here $\lambda(p)$ and $\eta(p)$ are real-valued functions.  Once we specify a cosmological background, $\lambda(p)$ is a fixed function (determined by the cosmological background, as described in the previous Subsection), but $\eta(p)$ is a free function (which may be chosen arbitrarily).

It is straightforward to check that, if the matrices $B_{\pm}(p)$ have this form, then the operators $\{a,b\}$ transform under $C$, $P$ and $T$ as follows:
\begin{equation}
  \label{CPT_scalar}
  \begin{array}{rclrcl}
    C a({\bf p})C^{-1}\!\!&\!=\!&\!\xi_{c}^{\ast} b(+{\bf p}), &\;\,
    C b_{}^{\dagger}({\bf p})C^{-1}\!\!&\!=\!&\!\xi_{c}^{\ast} a_{}^{\dagger}(+{\bf p}), \\
    P a({\bf p})P^{-1}\!\!&\!=\!&\!\xi_{p}^{\ast} a(-{\bf p}), &\;\,
    P b_{}^{\dagger}({\bf p})P^{-1}\!\!&\!=\!&\!\xi_{p}^{\ast} b_{}^{\dagger}(-{\bf p}), \\
    T a({\bf p})T^{-1}\!\!&\!=\!&\!\xi_{t}^{\ast} a(-{\bf p}), &\;\,
    T b_{}^{\dagger}({\bf p})T^{-1}\!\!&\!=\!&\!\xi_{t}^{\ast} b_{}^{\dagger}(-{\bf p}),
   \end{array}
\end{equation}
where $\xi_{c}$, $\xi_{p}$ and $\xi_{t}$ are the $C$, $P$ and $T$ phases introduced above in Subsection \ref{canon_quant_scalar}.  In what follows we can choose $\xi_{c}\xi_{p}\xi_{t}=1$ (see Eq. 5.8.4 in Ref. \cite{WeinbergQFT1}).  Note that the operators $C$, $P$ and $T$ commute, and that $C$ and $P$ are unitary, while $T$ is anti-unitary.  This means that $C$ and $P$ are linear, while $T$ is anti-linear; and, in all three cases, the adjoint of the operator is equal to its inverse. Thus, from the relations (\ref{CPT_scalar}) we may infer, for example, that $T b({\bf p})T^{-1}=\xi_{t}b(-{\bf p})$, etc.  

Thus, if we define $|0_{\eta}\rangle$ to be the state that is annihilated by all of the annihilation operators $a({\bf p})$ 
and $b({\bf p})$:
\begin{equation}
  a({\bf p})|0_{\eta}\rangle=b({\bf p})|0_{\eta}\rangle=0\quad(\forall\,{\bf p})
\end{equation}
the relations (\ref{CPT_scalar}) imply $[CPT]a({\bf p})[CPT]^{-1}\propto b({\bf p})$ and $[CPT]b({\bf p})[CPT]^{-1}\propto a({\bf p})$, so that this vacuum is, indeed, $CPT$-invariant:
\begin{equation}
  CPT|0_{\eta}\rangle\propto|0_{\eta}\rangle.
\end{equation} 
We have written the vacuum $|0_{\eta}\rangle$ with a subscript "$\eta$" to emphasize that each choice for the free function
$\eta(p)$ defines a different, inequivalent, $CPT$-invariant vacuum.  

This is similar to the situation in de Sitter space: when one looks for a vacuum that respects the full symmetry of de Sitter, one finds that the answer is not unique -- instead, there are a family of such vacua (the so-called $\alpha$-vacua \cite{Allen:1985ux}).  Similarly, in an FRW spacetime with time-reversal symmetry, when we look for a vacuum that respects the full symmetry of the background (and, in particular, that respects $CPT$) we find that the answer is not unique -- instead, there are a family of such vacua (the $\eta$-vacua constructed above).

We end this subsection by giving the $CPT$-invariant mode functions $u(p,\tau)$ explicitly.  In the special case where $\eta=0$, the corresponding $CPT$-invariant mode function $u_{0}(p,\tau)$ is neatly expressed in terms of the "in" and "out" mode functions $u_{-}$ and $u_{+}$ as follows:
\begin{equation}
  u_{0}(p,\tau)=\frac{1}{2\,{\rm cosh}[\lambda(p)/2]}[u_{+}(p,\tau)+u_{-}(p,\tau)].
\end{equation}
As we will see in Subsection \ref{CPT_vacuum_subsec_scalar}, $u_{0}(p,\tau)$ is the preferred $CPT$-invariant mode function.  The more general ($\eta\neq0$) $CPT$-invariant mode function $u(p,\tau)$ is then expressed in terms of the preferred ($\eta=0$) mode function $u_{0}(p,\tau)$ as follows:
\begin{equation}
  u(p,\tau)={\rm cosh}[\eta(p)]u_{0}(p,\tau)+{\rm sinh}[\eta(p)]u_{0}^{\ast}(p,\tau).
\end{equation}

\subsection{The preferred $CPT$ invariant vacuum}
\label{CPT_vacuum_subsec_scalar}       

In this subsection we show that, among the $CPT$-invariant vacua $|0_{\eta}\rangle$ on an time-reversal-symmetric FRW background, one particular vacuum is preferred: the $\eta=0$ vacuum $|0_{0}\rangle$.  This is again similar to the situation in de Sitter space where, among the de Sitter invariant vacua (the $\alpha$ vacua), one is preferred (the "Bunch-Davies" vacuum).

Here are two simple arguments that both lead us to the preferred vacuum $|0_{0}\rangle$:  
\begin{itemize}
\item First, consider the quantity $\langle 0_{\eta}|a_{\pm}^{\dagger}({\bf p})a_{\pm}^{}({\bf p})|0_{\eta}\rangle$ -- {\it i.e.}\ the expectation value of the number operator for particles of momentum ${\bf p}$, according to an asymptotic observer (long before or long after the bang), assuming that the universe is in the $CPT$-invariant vacuum state $|0_{\eta}\rangle$.  Using Eqs.~(\ref{Bogoliubov_pm_scalar}, \ref{B_pm_explicit_scalar}), we find
\begin{equation}
    \langle 0_{\eta}|a_{\pm}^{\dagger}({\bf p})a_{\pm}^{}({\bf p})|0_{\eta}\rangle
    =(2\pi)^{3}\delta({\bf 0})|\beta_{\pm}(p)|^{2} 
\end{equation}
where
\begin{equation}
  |\beta_{\pm}(p)|^{2}=\frac{{\rm cosh}[2\eta(p)]{\rm cosh}[\lambda(p)]-1}{2}.
\end{equation}
Dividing out the uninteresting divergence $\delta({\bf 0})$ coming from the infinite spatial volume, we see that on a given FRW background spacetime ({\it i.e.}\ for fixed $\lambda$), the number density of particles (according to an asymptotic observer) is minimized when $\eta(p)=0$.
\item Second, consider the quantity $\langle 0_{\eta}|H|0_{\eta}\rangle$ -- {\it i.e.}\ the expectation value of the Hamiltonian 
$H$, according to an asymptotic observer, assuming the universe is in the $CPT$-invariant vacuum state $|0_{\eta}\rangle$.
From (\ref{L_phi}) we first obtain the Hamiltonian
\begin{equation}
  \label{H_phi}
  H=\int d^{3}x\Big[(\varphi')^{\dagger}(\varphi')+(\nabla\varphi)^{\dagger}(\nabla\varphi)+\mu^{2}\varphi^{\dagger}\varphi\Big],
\end{equation}
and then, using Eqs.~(\ref{phi_expansion_2}, \ref{scalar_bc}, \ref{u_from_upm}, \ref{B_pm_explicit_scalar}), we find that asymptotically (as $\tau\to\pm\infty$) 
$\langle 0_{\eta}|H|0_{\eta}\rangle$ is given by
\begin{subequations}
  \begin{eqnarray}
    &&\delta({\bf 0})\int d^{\;\!3}p\;\omega(p,\tau)\Big[|\alpha_{\pm}(p)|^{2}+|\beta_{\pm}(p)|^{2}\Big] \\
    &=&\delta({\bf 0})\int d^{\;\!3}p\;\omega(p,\tau){\rm cosh}[2\eta(p)]{\rm cosh}[\lambda(p)]\,.
  \end{eqnarray}
\end{subequations}
Once again, after dividing out $\delta({\bf 0})$, we see that on a given FRW background spacetime ({\it i.e.}\ 
for fixed $\lambda$), the energy density (according to an asymptotic observer) is minimized when $\eta(p)=0$.
\end{itemize}
Thus, among the $CPT$-invariant vacua $|0_{\eta}\rangle$, the vacuum $|0_{0}\rangle$ is the one that is least excited (in the sense of minimum expected particle density and energy density).  


\section{$CPT$ invariant vacuum in FRW: spinor field}

Now let us do the analogous analysis for spinors: the analysis closely parallels the preceding one, but there are a few important differences which crop up.

\subsection{The in/out bases}
\label{mode_expansion_spinor}

We consider a Dirac spinor field $\Psi$ with mass $m>0$ on a flat FRW background spacetime.  (The case of a Majorana spinor field can be handled by a straightforward specialization of the following analysis.)  The Lagrangian is
\begin{subequations}
  \label{Dirac_L}
  \begin{eqnarray} 
    L&=&\sqrt{-g}[i\bar{\Psi}e^{\mu}_{a}\gamma^{a}\nabla_{\mu}\Psi-m\bar{\Psi}\Psi] \\
    &=&i\bar{\psi}\partial\!\!\!\!\;\!/\,\psi-\mu\bar{\psi}\psi.
  \end{eqnarray}
\end{subequations}
On the first line of (\ref{Dirac_L}), we have the usual curved space Dirac operator with Levi-Civita connection; in comoving/conformal coordinates, the tetrad is $e^{\mu}_{a}=(1/a)\delta^{\mu}_{a}$, and $\gamma^{a}$ are the $4\times4$ Dirac gamma matrices which, in a standard basis (the Weyl basis), may be written in terms of the $2\times2$ Pauli matrices $\sigma^{i}$ as follows
\begin{subequations}
  \begin{eqnarray}
  \gamma^{0}&=&\left(\begin{array}{cc} \;\;0\;\; & \;\;1\;\; \\ 1 & 0 \end{array}\right),  \\
  \gamma^{i}&=&\left(\begin{array}{cc} 0 & \sigma^{i} \\ -\sigma^{i} & \;\;0\;\; \end{array}\right), \\
  \gamma^{5}&=&\left(\!\begin{array}{cc} \;-1\; & 0 \\ 0 & \;+1\;\end{array}\!\right).
  \end{eqnarray}
\end{subequations}
For convenience, on the second line of (\ref{Dirac_L}), we have introduced the Weyl invariant spinor field $\psi=\psi({\bf x},\tau)$ and its effective mass $\mu=\mu(\tau)$:
\begin{equation}
  \label{mu_spinor}
  \psi\equiv a^{3/2}\Psi\quad{\rm and}\quad \mu\equiv am,
\end{equation}
and $\partial\!\!\!\!\,/=\gamma^{\mu}\partial_{\mu}$ is the usual flat-space Dirac operator, where the partial derivatives $\partial_{\mu}$ are with respect to the comoving/conformal coordinates $\{{\bf x},\tau\}$.

From here we obtain the equation of motion
\begin{equation}
  \label{Dirac_eq}
  (i\partial\!\!\!\!\;\!/\,-\mu)\psi=0.
\end{equation}
Because of the FRW symmetry of the background, if $\psi({\bf x},\tau)$ is a solution of (\ref{Dirac_eq}), so is:
\begin{subequations}
  \begin{itemize}
    \item its spatial translation 
      \begin{equation}
        \psi_{{\bf y}}({\bf x},\tau)\equiv\psi({\bf x}+{\bf y},\tau),\;\;
      \end{equation}
    \item its spatial rotation
      \begin{equation}
        \psi_{R}({\bf x},\tau)\equiv\psi(R{\bf x},\tau),\;\;\;\;\;\;\;
      \end{equation}
    \item its charge conjugate
      \begin{equation}
        \psi_{c}({\bf x},\tau)\equiv-i\gamma^{2}\psi^{\ast}({\bf x},\tau),
      \end{equation}
    \item its parity reverse
      \begin{equation}
        \psi_{p}({\bf x},\tau)\equiv\gamma^{0}\psi(-{\bf x},\tau),\;\;\;\,
      \end{equation}
    \item and (if $a$ and hence $\mu$ is an even or odd function of $\tau$) its time reverse
      \begin{equation}
        \psi_{t}({\bf x},\tau)\equiv\left\{\begin{array}{rc}
        \gamma^{5}\gamma^{0}\psi_{c}({\bf x},-\tau) & \quad(\mu\;{\rm even}) \\
        \gamma^{0}\psi_{c}({\bf x},-\tau) & \quad(\mu\;\,{\rm odd}\,) \end{array}\right. .
      \end{equation}
  \end{itemize}
\end{subequations}

We emphasize that the spinor case has an important new wrinkle, compared to the scalar case.  In the scalar case, it is the {\it squared} mass, $\mu^{2}(\tau)$, that appears in the Lagrangian and the equation of motion -- so, for $T$ symmetry, the only relevant case is when $\mu^{2}$ is an even function of $\tau$.  But in the spinor case, it is $\mu(\tau)$ itself that appears in (\ref{Dirac_eq}); and so there are two relevant cases to consider: when $\mu$ is an even or odd function of $\tau$.\footnote{At first glance, a negative fermion mass might sound pathological.  But in fact, when the spinor mass $m$ is a constant, only its absolute value is of physical significance: we can flip a negative mass $-m$ to a positive mass $m$ by a change of variables (a chiral transformation $\psi\to {\rm exp}[i\gamma^{5}\pi/2]\psi$).}

We would like to expand the field $\psi(x)$ in a basis of solutions of (\ref{Dirac_eq}).  Let us consider how the various symmetries of the FRW background effect this expansion:
\begin{itemize}
\item {\it Spatial translations.}  Because of the spatial translation invariance, we can take the solutions to be spatial fourier modes: 
$\psi({\bf x},\tau)=\psi({\bf p},\tau){\rm e}^{i{\bf p}{\bf x}}$.  
\item \sout{{\it Boosts}}.  In Minkowski space, we can boost into the ${\bf p}=0$ rest frame of a massive particle; this means the little group is $SO(3)$, and the internal spin states of a massive particle of momentum ${\bf p}$ are labelled by the eigenvalues of $J_{3}$ (the $\hat{{\bf z}}$ component of the spin) \cite{WeinbergQFT1}.  By contrast, in FRW there is a preferred spatial slicing and boosts are not a symmetry, so the most we can do is rotate ${\bf p}$ into the fiducial momentum ${\bf k}=p\,\hat{{\bf z}}$ ($p\equiv|{\bf p}|$); this means the little group is $SO(2)$, and the states of a massive particle of momentum ${\bf p}$ are labelled by the eigenvalue $h$ of the helicity operator $\hat{{\bf p}}\cdot{\bf J}$.  In this sense, a massive particle in FRW resembles a {\it massless} particle in Minkowski.
\item {\it Charge conjugation.}  Eq.~(\ref{Dirac_eq}) has four independent solutions per wavevector ${\bf p}$.  If $\psi({\bf p},h,x)=u({\bf p},h,\tau){\rm e}^{i{\bf p}{\bf x}}$ denotes a positive frequency solution of (\ref{Dirac_eq}) with momentum ${\bf p}$ and helicity $h$, then we can take the four solutions to be: the two positive frequency solutions $\psi({\bf p},h,x)$ (with $h=\pm1/2$), and the two corresponding negative frequency solutions $\psi_{c}({\bf p},h,x)=-i\gamma^{2}\psi^{\ast}({\bf p},h,x)$.  We then expand the field as follows
\begin{subequations}
  \begin{eqnarray}
    \label{psi_expansion_1}
    \psi(x)
    \!&\!=\!&\!\!\sum_{h}\!\int\!\!\!\frac{d^{3}p}{(2\pi)^{3}}\!\left[a({\bf p},h)\psi({\bf p},h,x)\right. \nonumber \\
    \!&\!\!&\!\qquad\qquad\;\;\,\left.+b_{}^{\dagger}({\bf p},h)\psi_{c}({\bf p},h,x)\right]\!, \\
    \label{psi_expansion_2}
    \!&\!=\!&\!\!\sum_{h}\!\int\!\!\!\frac{d^{3}p}{(2\pi)^{3}}\!\left[a({\bf p},h)u({\bf p},h,\tau){\rm e}^{+i{\bf p}{\bf x}}\right. \nonumber\\
    \!&\!\!&\!\qquad\qquad\;\;\,\left.+b_{}^{\dagger}\!({\bf p},h)v({\bf p},h,\tau){\rm e}^{-i{\bf p}{\bf x}}\right]\!,
  \end{eqnarray}
\end{subequations}
where we have introduced the notation
\begin{equation}
  v({\bf p},h,\tau)\equiv-i\gamma^{2}u^{\ast}({\bf p},h,\tau).
\end{equation}
\item \sout{{\it Time translations.}}  Again, the ingoing (early time) and outgoing (late time) observers will perform the mode expansion (\ref{psi_expansion_2}) in two different ways:
\begin{eqnarray}
  \label{psi_expansion_pm}
  \psi(x)\!&\!=\!&\!\!\sum_{h}\!\int\!\!\!\frac{d^{3}p}{(2\pi)^{3}}\!\!\left[a_{\pm}({\bf p},h)u_{\pm}({\bf p},h,\tau){\rm e}^{+i{\bf p}{\bf x}}\right. \nonumber\\
    \!&\!\!&\!\qquad\qquad\;\;\,\left.+b_{\pm}^{\dagger}({\bf p},h)v_{\pm}({\bf p},h,\tau){\rm e}^{-i{\bf p}{\bf x}}\right]\!.
\end{eqnarray}
The $-$ and $+$ subscripts correspond to the in and out observers, respectively, and $v_{\pm}({\bf p},h,\tau)\equiv-i\gamma^{2}u_{\pm}^{\ast}({\bf p},h,\tau)$.  The ingoing and outgoing bases are fixed as follows:

\item {\it Spatial Rotations}.  First consider the fiducial momentum ${\bf k}=p\,\hat{{\bf z}}$.  The corresponding positive frequency modes $u_{\pm}(p\,\hat{{\bf z}},h,\tau)$ are fixed by the boundary conditions
\begin{equation}
  \label{spinor_bc}
    u_{\pm}(p\,\hat{{\bf z}},h,\tau)\!\to\!\hat{u}(p,h,\tau)\;\!{\rm exp}\!\left[\!-i\!\!\int_{\pm\tau_{0}^{}(p)}^{\tau}\!\!\!\!\!\!\!\!\!\!\!\!\!\omega(p,\tilde{\tau})d\tilde{\tau}\right]\!\!\!\!
\end{equation}
as $\tau\to\pm\infty$, where $\omega(p,\tau)>0$ is the positive root of $\omega^{2}\equiv p^{2}+\mu^{2}$, $\tau_{0}^{}(p)$ is arbitrary and may be fixed for convenience (we return to this in Subsection \ref{spinor_bogoliubov_section}), $\hat{u}(p,1/2,\tau)$ is given by
\begin{subequations}
  \label{uhat}
  \begin{equation}
    \frac{1}{\sqrt{\mu^{2}(\tau)\!+\!(\omega(p,\tau)\!+\!p)^{2}}}
    \!\!\left[\!\!\begin{array}{c} \mu(\tau) \\ 0 \\ \omega(p,\tau)\!+\!p \\ 0 \end{array}\!\!\right]\!,
  \end{equation}
  and $\hat{u}(p,-1/2,\tau)$ is given by
  \begin{equation}
    \frac{1}{\sqrt{\mu^{2}(\tau)\!+\!(\omega(p,\tau)\!+\!p)^{2}}}
    \!\!\left[\!\!\begin{array}{c} 0 \\ \omega(p,\tau)\!+\!p \\ 0 \\ \mu(\tau) \end{array}\!\!\right]\!.
  \end{equation}
\end{subequations}
Next consider the arbitrary momentum ${\bf p}$: the corresponding positive frequency solutions $u_{\pm}({\bf p},h,\tau)$ are obtained from the fiducial solutions $u_{\pm}(p\,\hat{{\bf z}},h,\tau)$ by applying an appropriate rotation $U[R(\hat{{\bf p}})]$ \cite{WeinbergQFT1}
\begin{subequations}
  \label{general_uv_from_fiducial_uv}
  \begin{eqnarray}
    \label{general_u_from_fiducial_u}
    u_{\pm}({\bf p},h,\tau)&=&U[R(\hat{{\bf p}})]u_{\pm}(p\,\hat{{\bf z}},h,\tau) \\
    \label{general_v_from_fiducial_v}
    \Rightarrow\quad
    v_{\;\!\pm}({\bf p},h,\tau)&=&U[R(\hat{{\bf p}})]v_{\;\!\pm}(p\,\hat{{\bf z}},h,\tau) 
  \end{eqnarray}
\end{subequations}
In particular, if the unit vector $\hat{{\bf p}}\equiv{\bf p}/|{\bf p}|$ points in the direction characterized by spherical coordinates $\{\theta,\phi\}$:
\begin{equation}
  \hat{{\bf p}}=\{{\rm sin}(\theta){\rm cos}(\phi),{\rm sin}(\theta){\rm sin}(\phi),{\rm cos}(\theta)\}
\end{equation}
with $0\leq\theta\leq\pi$ and $-\pi\leq\phi<\pi$, then the rotation $U[R(\hat{{\bf p}})]$ is
\begin{equation}
  \label{U(R)}
  U[R(\hat{{\bf p}})]={\rm exp}\left[-i\phi {\bf J}_{3}\right]{\rm exp}\left[-i\theta {\bf J}_{2}\right]
\end{equation}
where
\begin{equation}
  {\bf J}_{k}=\frac{1}{2}\left(\begin{array}{cc} \sigma_{k} & 0 \\ 0 & \sigma_{k} \end{array}\right).
\end{equation}

In this way, all the ingoing and outgoing basis modes, $u_{\pm}({\bf p},h,\tau)$ and $v_{\pm}({\bf p},h,\tau)$, are fixed.  One can check that these modes are eigenvectors of the helicity operator $\hat{{\bf p}}\cdot{\bf J}$:
\begin{subequations}
  \label{helicity_pm}
  \begin{eqnarray}
    (\hat{{\bf p}}\cdot{\bf J})u_{\pm}({\bf p},h,\tau)&=&+h\,u_{\pm}({\bf p},h,\tau), \\
    (\hat{{\bf p}}\cdot{\bf J})v_{\pm}({\bf p},h,\tau)&=&-h\,v_{\pm}({\bf p},h,\tau).
  \end{eqnarray}
\end{subequations}

\item {\it Parity.} If $\psi(\tau,{\bf x})=u_{\pm}({\bf p},h,\tau){\rm e}^{i{\bf p}{\bf x}}$ is an outgoing (ingoing) positive frequency solution with momentum ${\bf p}$ and helicity $h$, then $\psi_{p}(\tau,{\bf x})$ is an outgoing (ingoing) positive frequency solution with momentum $-{\bf p}$ and helicity $-h$: $\propto u_{\pm}(-{\bf p},-h,\tau){\rm e}^{i(-{\bf p}){\bf x}}$.  So $\gamma^{0}u_{\pm}({\bf p},h,\tau)\propto u_{\pm}(-{\bf p},-h,\tau)$ and, in particular, Eqs.~(\ref{spinor_bc}, \ref{uhat}, \ref{general_uv_from_fiducial_uv}, \ref{U(R)}) imply
\begin{equation}
  \label{upm_P_spinor}
  u_{\pm}(-{\bf p},-h,\tau)=-i\,{\rm sgn}(\phi)\,\gamma^{0}u_{\pm}({\bf p},h,\tau).
\end{equation}
Here we have used the fact that, if ${\bf p}$ is characterized by spherical coordinates $\{\theta,\phi\}$ then $-{\bf p}$ is 
characterized by spherical coordinates $\{\tilde{\theta},\tilde{\phi}\}=\{\pi-\theta,\phi-{\rm sgn}(\phi)\pi\}$.

\item {\it Time reversal.}  If $\mu$ is an even or odd function of $\tau$, and $\psi(\tau,{\bf x})=u_{\pm}({\bf p},h,\tau){\rm e}^{i{\bf p}{\bf x}}$ is an outgoing (ingoing) positive frequency solution with momentum ${\bf p}$ and helicity $h$, then $\psi_{t}(\tau,{\bf x})$ is an ingoing (outgoing) positive frequency solution with momentum $-{\bf p}$ and helicity $h$: $\propto u_{\mp}(-{\bf p},h,\tau){\rm e}^{i(-{\bf p}){\bf x}}$.  Thus, for $\mu(\tau)$ even or odd, respectively, either $\gamma^{5}\gamma^{0}v_{\pm}({\bf p},h,-\tau)$ or $\gamma^{0}v_{\pm}({\bf p},h,-\tau)$ is $\propto u_{\mp}(-{\bf p},h,\tau)$ and, in particular, Eqs.~(\ref{spinor_bc}, \ref{uhat}, \ref{general_uv_from_fiducial_uv}, \ref{U(R)}) imply
\begin{equation}
  \label{upm_T_spinor}
  u_{\mp}(-{\bf p},h,\tau)\!=\!\left\{\!\!\!\begin{array}{c}
  (2h)\gamma^{5} \\ 1 \end{array}\!\!\right\}\!
  \frac{\gamma^{0}v_{\pm}({\bf p},h,-\tau)}{i\,{\rm sgn}(\phi)} 
  \;\;\;\left\{\!\!\begin{array}{c} \mu\;{\rm even} \\ \mu\;{\rm odd} \end{array}\!\!\right\}\!.
\end{equation}
\end{itemize}

\subsection{Canonical quantization in the in/out bases}
\label{canon_quant_spinor}

From (\ref{Dirac_L}) we obtain the conjugate momentum 
\begin{equation}
  \pi=\frac{\partial L}{\partial \psi'}=i\psi^{\dagger}.
\end{equation}
With the expansion (\ref{psi_expansion_pm}) the canonical anti-commutation relations
\begin{equation}
  \label{canonical_comm_spinor}
  \{\psi({\bf x},\tau),\pi({\bf y},\tau)\}=i\delta({\bf x}-{\bf y})\mathbb{I}_{\;\!4\times4}\quad({\rm all}\;{\rm others}\;{\rm vanish})
\end{equation}
are equivalent to the usual anti-commutation relations for fermionic creation and annihilation operators
\begin{eqnarray}
  \label{ab_comm_pm_spinor}
  (2\pi)^{3}\delta({\bf p}-{\bf q})\delta_{h,h_{\ast}}\!&\!=\!&\!
  \{a_{\pm}({\bf p},h),a_{\pm}^{\dagger}({\bf q},h_{\ast})\} \nonumber \\
  \!&\!=\!&\!
  \{b_{\pm}({\bf p},h),b_{\pm}^{\dagger}({\bf q},h_{\ast})\}\quad
\end{eqnarray}
(all others vanish), provided the mode functions satisfy the normalization condition
\begin{eqnarray}
  \label{spinor_normalization_pm}
  \mathbb{I}_{\;\!4\times4}\!&\!=\!&\!
  \sum_{h}[u_{\pm}^{}({\bf p},h,\tau)u_{\pm}^{\ast}({\bf p},h,\tau) \nonumber\\
  \!&\!\!&\!\quad+v_{\pm}^{}(-{\bf p},h,\tau)v_{\pm}^{\ast}(-{\bf p},h,\tau)].
\end{eqnarray}
The interpretation is that $a^{\dagger}({\bf p},h)$ creates a particle with momentum ${\bf p}$ and helicity $h$, while $b^{\dagger}({\bf p},h)$ creates the corresponding anti-particle with momentum ${\bf p}$ and helicity $h$.

Now let us see how these operators transform under $C$, $P$ and $T$ (see \cite{WeinbergQFT1}):
\begin{itemize}
\item $C$:  The requirement that $\psi$ should transform under charge conjugation like
\begin{equation}
  C\psi(x)C^{-1}=\xi_{c}^{\ast}\psi_{c}(x),
\end{equation}
(where $C$ is unitary and $\xi_{c}$ is the associated charge conjugation phase) implies that the creation and annihilation operators in (\ref{psi_expansion_pm}) transform like
\begin{subequations}
  \begin{eqnarray}
    Ca_{\pm}^{}({\bf p},h)C^{-1}&=&\xi_{c}^{\ast}b_{\pm}^{}({\bf p},h), \\
    Cb_{\pm}^{\dagger}({\bf p},h)C^{-1}&=&\xi_{c}^{\ast}a_{\pm}^{\dagger}({\bf p},h).
  \end{eqnarray}
\end{subequations}

\item $P$: The requirement that $\psi$ should transform under parity like 
\begin{equation}
  P\psi(x)P^{-1}=\xi_{p}^{\ast}\psi_{p}(x),
\end{equation}
(where $P$ is unitary and $\xi_{p}$ is the associated parity phase) implies that the creation and annihilation operators in (\ref{psi_expansion_pm}) transform like
\begin{subequations}
  \begin{eqnarray}
    Pa_{\pm}^{}({\bf p},h)P^{-1}&=&-i\xi_{p}^{\ast}{\rm sgn}(\phi)\,a_{\pm}^{}(-{\bf p},-h), \\
    Pb_{\;\!\pm}^{\dagger}({\bf p},h)P^{-1}&=&-i\xi_{p}^{\ast}{\rm sgn}(\phi)\,b_{\;\!\pm}^{\dagger}(-{\bf p},-h).
  \end{eqnarray}
\end{subequations}

\item $T$:  The requirement that $\psi$ should transform under time reversal like 
\begin{equation}
  T\psi(x)T^{-1}=\xi_{t}^{\ast}\psi_{t}^{\ast}(x),
\end{equation}
(where $T$ is anti-unitary and $\xi_{t}$ is the associated time-reversal phase), implies that the creation and annihilation operators in (\ref{psi_expansion_pm}) transform like
\begin{subequations}
  \begin{equation}
    T a_{\pm}^{}({\bf p},h)T^{-1}\!=\!\left\{\!\!\!\begin{array}{c} +2h \\ 1 \end{array}\!\!\!\right\}
    i\xi_{t}^{\ast}{\rm sgn}(\phi)a_{\mp}^{}(\!-{\bf p},h)\quad
    \left\{\!\!\begin{array}{c} \mu\;{\rm even} \\ \mu\;\,{\rm odd}\, \end{array}\!\!\right\}\!,
  \end{equation}
  \begin{equation}
    T b_{\;\!\pm}^{\dagger}({\bf p},h)T^{-1}\!=\!\left\{\!\!\!\begin{array}{c} -2h \\ 1 \end{array}\!\!\!\right\}
    i\xi_{t}^{\ast}{\rm sgn}(\phi)b_{\;\!\mp}^{\dagger}(\!-{\bf p},h)\quad
    \left\{\!\!\begin{array}{c} \mu\;{\rm even} \\ \mu\;\,{\rm odd}\, \end{array}\!\!\right\}\!.
  \end{equation}
\end{subequations}
\end{itemize}

\subsection{Transformation between the in/out bases}
\label{spinor_bogoliubov_section}

The spatial translational and rotational symmetry of the background imply that an ingoing positive frequency solution $u_{-}({\bf p},h,\tau)$ must evolve into a linear combination of the outgoing positive and negative frequency solutions $u_{+}({\bf p},h,\tau)$ and $v_{+}(-{\bf p},h,\tau)$, with some coefficients $\alpha({\bf p},h)$ and $\beta({\bf p},h)$:
\begin{equation}
  \label{um_from_up_spinor}
  u_{-}({\bf p},h,\tau)=\alpha({\bf p},h)u_{+}({\bf p},h,\tau)+\beta({\bf p},h)v_{+}(-{\bf p},h,\tau).
\end{equation}
Eqs.~(\ref{general_uv_from_fiducial_uv}, \ref{U(R)}) imply that $\alpha({\bf p},h)$ is independent of $\hat{{\bf p}}$: $\alpha({\bf p},h)=\alpha(p,h)$; while $\beta({\bf p},h)$ is almost independent of $\hat{{\bf p}}$: $\beta({\bf p},h)={\rm sgn}(\phi)\hat{\beta}(p,h)$.  The parity condition (\ref{upm_P_spinor}) then implies that $\alpha({\bf p},h)$ and $\beta({\bf p},h)$ are also independent of $h$: 
$\alpha({\bf p},h)=\alpha(p)$ and $\beta({\bf p},h)={\rm sgn}(\phi)\hat{\beta}(p)$.  And the time-reversal condition (\ref{upm_T_spinor}) further implies that $|\alpha(p)|^{2}+|\hat{\beta}(p)|^{2}=1$ and that $\hat{\beta}(p)$ is real or imaginary, when $\mu$ is even or odd, respectively.  Then, adjusting the choice of $\tau_{0}^{}(p)$ in (\ref{spinor_bc}) adjusts $u_{\pm}({\bf p},h,\tau)\to{\rm e}^{\pm i\chi(p)}u_{\pm}({\bf p},h,\tau)$ and $v_{\pm}({\bf p},h,\tau)\to{\rm e}^{\mp i\chi(p)}v_{\pm}({\bf p},h,\tau)$, and we use this freedom to phase-rotate $\alpha(p)$ to make it real and non-negative.  Putting this all together, we can write 
\begin{equation}
  \alpha({\bf p},h)={\rm cos}[\lambda({\bf p})]\;\;{\rm and}\;\;
  \beta({\bf p},h)=\kappa^{1/2}{\rm sin}[\lambda({\bf p})]
\end{equation}
where $\lambda({\bf p})={\rm sgn}(\phi)\hat{\lambda}(p)$ is a real function with $-\pi/2<\hat{\lambda}(p)<\pi/2$, and the $\pm$ sign $\kappa$ denotes the parity of $\mu$: $\mu(-\tau)=\kappa\mu(\tau)$.

If we substitute (\ref{um_from_up_spinor}) into (\ref{psi_expansion_pm}), we see that the Bogoliubov transformation 
\begin{equation}
  \label{spinor_bogoliubov}
  \left[\begin{array}{c} a_{+}^{}(+{\bf p},h) \\ b_{\;\!+}^{\dagger}(-{\bf p},h) \end{array}\right]=B({\bf p})
   \left[\begin{array}{c} a_{-}^{}(+{\bf p},h) \\ b_{\;\!-}^{\dagger}(-{\bf p},h)\end{array}\right]
\end{equation}
between the "in" operators ($a_{-}$ and $b_{-}$) and the "out" operators ($a_{+}$ and $b_{+}$) is encoded in the 
Bogoliubov matrix $B$
\begin{equation}
  \label{B_spinor}
    B({\bf p})\equiv \left[\begin{array}{rr} {\rm cos}[\lambda({\bf p})]\qquad & \frac{-1}{\kappa_{}^{1/2}}{\rm sin}[\lambda({\bf p})] \\
    \kappa_{}^{1/2}{\rm sin}[\lambda({\bf p})]\qquad & {\rm cos}[\lambda({\bf p})] \end{array}\right].
\end{equation}

Again, note that the transformation (\ref{spinor_bogoliubov}) makes physical sense.  The (FRW) background does not break internal $U(1)$ invariance, spatial translational invariance, or the little group $SO(2)$ corresponding to rotations around ${\bf p}$.  Hence, the Bogoliubov transformation should "conserve" charge, spatial momentum, and spin (around the ${\bf p}$ axis): {\it i.e.}\ the "out" operator $a_{+}({\bf p},h)$ (which annihilates a particle of momentum ${\bf p}$ and spin $h$ in the ${\bf p}$ direction) can only be a linear combination of the "in" operator $a_{-}({\bf p},h)$ (which also annihilates a particle of momentum ${\bf p}$ and spin $h$ in the ${\bf p}$ direction) and the "in" operator $b_{-}^{\dagger}(-{\bf p},h)$ (which creates an anti-particle of momentum $-{\bf p}$ and spin $-h$ in the ${\bf p}$ direction).

As in the scalar case, the function $|\lambda({\bf p})|$ measures the physical offset between the "in" vacuum $|0_{-}\rangle$ (the state annihilated by all the $a_{-}$ and $b_{-}$) and the "out" vacuum $|0_{+}\rangle$ (the state annihilated by all the $a_{+}$ and $b_{+}$):
\begin{subequations}
  \begin{eqnarray}
    a_{-}({\bf p},h)|0_{-}\rangle=b_{-}({\bf p},h)|0_{-}\rangle&=&0, \\
    a_{+}({\bf p},h)|0_{+}\rangle=b_{+}({\bf p},h)|0_{+}\rangle&=&0.
  \end{eqnarray}
\end{subequations}
And, as in the scalar case: if $\lambda({\bf p})$ is non-vanishing, the "in" and "out" vacua are inequivalent; and, in this case, even if the background is invariant under $CPT$, the corresponding in and out vacua are not:
\begin{subequations}
  \begin{eqnarray}
    CPT|0_{-}\rangle&\propto&|0_{+}\rangle, \\
    CPT|0_{+}\rangle&\propto&|0_{-}\rangle.  
  \end{eqnarray}
\end{subequations}

\subsection{CPT invariant bases and vacua}
\label{CPT_vacua_spinor}

In this subsection, we construct all the vacua that {\it are} $CPT$ invariant.

As in the scalar case, it was the requirement (\ref{spinor_bc}) that the basis modes $u$ have purely positive frequency (in either the far past or the far future) that forced us to introduce two inequivalent bases ("in" and "out"), and hence two inequivalent vacua ($|0_{-}\rangle$ and $|0_{+}\rangle$), that are swapped by $CPT$.  To construct bases and vacua that are {\it preserved} by $CPT$, we
must give up this requirement.  We replace the two expansions (\ref{psi_expansion_pm}) by the single expansion (\ref{psi_expansion_2}), while the conditions (\ref{general_uv_from_fiducial_uv}), (\ref{helicity_pm}), (\ref{upm_P_spinor}) and (\ref{upm_T_spinor}) are replaced by  
\begin{equation}
  \label{general_uv_from_fiducial_uv_0}
  \begin{array}{rcl}
    u({\bf p},h,\tau)&=&U[R(\hat{{\bf p}})]u(p\,\hat{{\bf z}},h,\tau), \\
    v({\bf p},h,\tau)&=&U[R(\hat{{\bf p}})]v(p\,\hat{{\bf z}},h,\tau), 
  \end{array}
\end{equation}
\begin{equation}
  \label{helicity_0}
  \begin{array}{rcl}
    (\hat{{\bf p}}\cdot{\bf J})u({\bf p},h,\tau)&=&+h\,u({\bf p},h,\tau), \\
    (\hat{{\bf p}}\cdot{\bf J})v({\bf p},h,\tau)&=&-h\,v({\bf p},h,\tau),
  \end{array}
\end{equation}
\begin{equation}
  \label{u0_P_spinor}
  u(-{\bf p},-h,\tau)=-i\,{\rm sgn}(\phi)\,\gamma^{0}u({\bf p},h,\tau),
\end{equation}
\begin{equation}
  \label{u0_T_spinor}
  u(-{\bf p},h,\tau)\!=\!\left\{\!\!\begin{array}{c}
  (2h)\gamma^{5} \\ 1 \end{array}\!\!\right\}\!\frac{\gamma^{0}v({\bf p},h,-\tau)}{i\,{\rm sgn}(\phi)} 
  \quad\left\{\!\!\begin{array}{c} \mu\;{\rm even} \\ \mu\;\,{\rm odd} \end{array}\!\!\right\}\!.
\end{equation}

Once again, the canonical equal-time anti-commutation relations (\ref{canonical_comm_spinor}) are equivalent to the 
usual anti-commutation relations for creation and annihilation operators
\begin{eqnarray}
  \label{ab_comm_0_spinor}
  (2\pi)^{3}\delta({\bf p}-{\bf q})\delta_{h,h_{\ast}}&=&\{a({\bf p},h),a^{\dagger}({\bf q},h_{\ast})\} \nonumber\\
  &=&\{b({\bf p},h),b^{\dagger}({\bf q},h_{\ast})\}
\end{eqnarray}
provided the mode functions satisfy the Wronskian normalization condition 
\begin{equation}
  \label{spinor_normalization_0}
  \sum_{h}[u({\bf p},h,\tau)u^{\ast}({\bf p},h,\tau)
  +v(-{\bf p},h,\tau)v^{\ast}(-{\bf p},h,\tau)]=\mathbb{I}_{\;\!4\times4}.
\end{equation}

Now, we can re-express the solution $u({\bf p},h,\tau){\rm e}^{i{\bf p}{\bf x}}$ as a linear combination of either the "in" solutions, $u_{-}({\bf p},h,\tau){\rm e}^{i{\bf p}{\bf x}}$ and $v_{-}(-{\bf p},h,\tau){\rm e}^{i{\bf p}{\bf x}}$, or the "out" solutions, $u_{+}({\bf p},h,\tau){\rm e}^{i{\bf p}{\bf x}}$ and $v_{+}(-{\bf p},h,\tau){\rm e}^{i{\bf p}{\bf x}}$:
\begin{equation}
  \label{u_from_upm_spinor}
  u({\bf p},h,\tau)=\alpha_{\pm}({\bf p},h)u_{\pm}^{}({\bf p},h,\tau)+\beta_{\pm}({\bf p},h)v_{\pm}^{}(-{\bf p},h,\tau).
\end{equation}
Eqs.~(\ref{general_uv_from_fiducial_uv}, \ref{U(R)}, \ref{general_uv_from_fiducial_uv_0}) imply that $\alpha_{\pm}({\bf p},h)$ is independent of $\hat{{\bf p}}$: $\alpha_{\pm}({\bf p},h)=\alpha_{\pm}(p,h)$; while $\beta_{\pm}({\bf p},h)$ is almost independent of $\hat{{\bf p}}$: $\beta_{\pm}({\bf p},h)={\rm sgn}(\phi)\hat{\beta}_{\pm}(p,h)$.  The parity conditions (\ref{upm_P_spinor}, \ref{u0_P_spinor}) then imply that $\alpha_{\pm}({\bf p},h)$ and $\beta_{\pm}({\bf p},h)$ are also independent of $h$: $\alpha_{\pm}({\bf p},h)=\alpha_{\pm}(p)$ and $\beta_{\pm}({\bf p},h)={\rm sgn}(\phi)\hat{\beta}_{\pm}(p)$.

If we now substitute (\ref{u_from_upm_spinor}) into (\ref{psi_expansion_2}) and compare to (\ref{psi_expansion_pm}), we see that the Bogoliubov transformations
\begin{equation}
  \label{Bogoliubov_pm_spinor}
  \left[\begin{array}{c} a_{\pm}^{}(+{\bf p},h) \\ b_{\pm}^{\dagger}(-{\bf p},h) \end{array}\right]=
  B_{\pm}({\bf p})
  \left[\begin{array}{c} a\,(+{\bf p},h) \\ b_{}^{\dagger\!}(-{\bf p},h) \end{array}\right]
\end{equation}
from the operators $\{a,b\}$ to the "in" or "out" operators, $\{a_{-},b_{-}\}$ or $\{a_{+},b_{+}\}$, are described, respectively, by the Bogoliubov matrices $B_{-}$ and $B_{+}$:
\begin{subequations}
  \label{B_pm_spinor}
  \begin{eqnarray}
  B_{\pm}({\bf p})&\equiv&\left[\begin{array}{cc} 
  \alpha_{\pm}^{}({\bf p}) & \quad\beta_{\pm}^{\ast}(-{\bf p}) \\
  \beta_{\pm}^{}({\bf p}) & \quad\alpha_{\pm}^{\ast}(-{\bf p}) \end{array}\right] \\ 
  &=&\left[\begin{array}{cc} \alpha_{\pm}^{}(p) & \quad-{\rm sgn}(\phi)\hat{\beta}_{\pm}^{\ast}(p) \\
  +{\rm sgn}(\phi)\hat{\beta}_{\pm}^{}(p) & \quad\alpha_{\pm}^{\ast}(p)\end{array}\right].\qquad
\end{eqnarray}
\end{subequations}
Note that $B_{+}$ and $B_{-}$ must obey the following three constraints:
\begin{enumerate}
\begin{subequations}
\item First, compatibility between the anti-commutation relations (\ref{ab_comm_pm_spinor}) and (\ref{ab_comm_0_spinor}) implies
\begin{equation}
  \label{alpha_beta_pm_normalization_spinor}
  {\rm Det}[B_{\pm}({\bf p})]=1.
\end{equation}
\item Second, compatibility between the time-reversal constraints (\ref{upm_T_spinor}) and (\ref{u0_T_spinor}) implies
\begin{equation}
  \label{alpha_beta_pm_T_spinor}
  B_{-}({\bf p})=B_{+}^{\ast}(-\kappa{\bf p})
\end{equation}
where, again, the $\pm$ sign $\kappa$ denotes the parity of $\mu$: $\mu(-\tau)=\kappa\mu(\tau)$.
\item Third, the map $B({\bf p})$ in Eq.~(\ref{B_spinor}), from the "in" operators $\{a_{-},b_{-}\}$ to the "out" operators $\{a_{+},b_{+}\}$, can be re-expressed in terms of $B_{+}$ and $B_{-}$ as follows
\begin{equation}
  B({\bf p})=B_{+}({\bf p})B_{-}^{-1}({\bf p}).
\end{equation}
\end{subequations}
\end{enumerate}
The most general solution for $B_{\pm}$ satisfying these three constraints is
\begin{equation}
  \label{B_pm_explicit_spinor}
  B_{\pm}({\bf p})=\hat{B}_{\pm}({\bf p})B_{0}({\bf p})
\end{equation}
where
\begin{subequations}
  \begin{eqnarray}
    \hat{B}_{\pm}({\bf p})\!&\!\equiv\!&\!\left[\!\begin{array}{rr} 
  {\rm cos}\frac{\lambda({\bf p})}{2} & \quad\frac{-1}{\pm(\kappa^{1/2})}{\rm sin}\frac{\lambda({\bf p})}{2} \\
  \pm(\kappa^{1/2}){\rm sin}\frac{\lambda({\bf p})}{2} & {\rm cos}\frac{\lambda({\bf p})}{2} \end{array}\!\right] \\
  B_{0}({\bf p})\!&\!\equiv\!&\!\left[\!\begin{array}{rr} 
  {\rm cos}\,\eta({\bf p}) & \quad\frac{-1}{(-\kappa)^{1/2}}{\rm sin}\,\eta({\bf p}) \\ (-\kappa)^{1/2}{\rm sin}\,\eta({\bf p}) & {\rm cos}\,\eta({\bf p}) \end{array}\!\right]\quad
  \end{eqnarray}
\end{subequations}
where $\lambda({\bf p})={\rm sgn}(\phi)\hat{\lambda}(p)$ and $\eta({\bf p})={\rm sgn}(\phi)\hat{\eta}(p)$ are real-valued functions.  Once we specify the cosmological background, $\hat{\lambda}(p)$ is a fixed function (determined by the cosmological background, as explained in Subsection \ref{spinor_bogoliubov_section}), but $\hat{\eta}(p)$ is a free function (which may be chosen arbitrarily).

It is straightforward to check that, if the matrices $B_{\pm}$ have this form, then the operators $\{a,b\}$ transform under $C$, $P$ and $T$ as follows:
\begin{equation}
  \begin{array}{rcl}
    \label{CPT_spinor_relations}
    C a\,({\bf p},h)C^{-1}\!&\!=\!&\!\xi_{c}^{\ast}\,b\,(+{\bf p},h), \\
    C b^{\dagger\!}({\bf p},h)C^{-1}\!&\!=\!&\!\xi_{c}^{\ast} a^{\dagger\!}(+{\bf p},h), \\
    P a\,({\bf p},h)P^{-1}\!&\!=\!&\!-i \xi_{p}^{\ast}{\rm sgn}(\phi)a\,(-{\bf p},-h), \\
    P b^{\dagger\!}({\bf p},h)P^{-1}\!&\!=\!&\!-i\xi_{p}^{\ast}{\rm sgn}(\phi)b^{\dagger\!}(-{\bf p},-h), \\
    T a({\bf p},h)\,T^{-1}\!&\!=\!&\!\left\{\!\!\!\begin{array}{c} +2h \\ 1 \end{array}\!\!\right\}
    i\xi_{t}^{\ast}{\rm sgn}(\phi)a\,(\!-{\bf p},h)
    \quad\left\{\!\!\begin{array}{c} \mu\;{\rm even} \\ \mu\;\,{\rm odd}\, \end{array}\!\!\right\}, \\ 
    T b^{\dagger\!}({\bf p},h)T^{-1}\!&\!=\!&\!\left\{\!\!\!\begin{array}{c} -2h \\ 1 \end{array}\!\!\right\}
    i\xi_{t}^{\ast}{\rm sgn}(\phi)b^{\dagger\!}(\!-{\bf p},h)
    \quad\left\{\!\!\begin{array}{c} \mu\;{\rm even} \\ \mu\;\,{\rm odd}\, \end{array}\!\!\right\},
   \end{array}
 \end{equation}
 where $\xi_{c}$, $\xi_{p}$ and $\xi_{t}$ are the $C$, $P$ and $T$ phases introduced above in Subsection \ref{canon_quant_scalar}.
 
Just as in the scalar case (see the discussion following (\ref{CPT_scalar})), and also noting that $\lambda(-{\bf p})=-\lambda({\bf p})$ and $\eta(-{\bf p})=-\eta({\bf p})$ as explained just after (\ref{B_pm_explicit_spinor}) above, the relations in (\ref{CPT_spinor_relations}) imply that $[CPT]a({\bf p},h)[CPT]^{-1}=-a({\bf p},-h)$ and $[CPT]b({\bf p},h)[CPT]^{-1}=-b({\bf p},-h)$.  Thus if we define $|0_{\eta}\rangle$ to be the state that is annihilated by all of the annihilation operators $a({\bf p},h)$ and $b({\bf p},h)$:
\begin{equation}
  a({\bf p},h)|0_{\eta}\rangle=b({\bf p},h)|0_{\eta}\rangle=0\quad(\forall\, {\bf p},h)
\end{equation}
then we see that it is, indeed, $CPT$-invariant:
\begin{equation}
  CPT|0_{\eta}\rangle\propto|0_{\eta}\rangle.
\end{equation}    
Again, we have written the vacuum $|0_{\eta}\rangle$ with a subscript "$\eta$" to emphasize that each choice for the free function $\eta(p)$ defines a different, inequivalent, $CPT$-invariant vacuum.  And again, there is a close parallel between these $CPT$-invariant "$\eta$-vacua" (in FRW) and the "$\alpha$-vacua" (in de Sitter).

We end by giving the $CPT$-invariant mode functions $u({\bf p},h,\tau)$ explicitly.  In the special case where $\eta=0$, the corresponding $CPT$-invariant mode function $u_{0}({\bf p},h,\tau)$ is neatly expressed in terms of the "in" and "out" mode functions $u_{-}$ and $u_{+}$ as follows:
\begin{equation}
  u_{0}({\bf p},h,\tau)=\frac{1}{2\,{\rm cos}[\lambda({\bf p})/2]}[u_{+}({\bf p},h,\tau)+u_{-}({\bf p},h,\tau)].
\end{equation}
As we will see in Subsection \ref{CPT_vacuum_subsec_spinor}, $u_{0}({\bf p},h,\tau)$ is the preferred $CPT$-invariant mode function.  The more general ($\eta\neq0$) $CPT$-invariant mode function $u({\bf p},h,\tau)$ is then expressed in terms of the preferred ($\eta=0$) mode function $u_{0}({\bf p},h,\tau)$ as follows:
\begin{eqnarray}
  u({\bf p},h,\tau)&=&{\rm cos}[\eta({\bf p})]u_{0}(+{\bf p},h,\tau) \nonumber\\
  &+&\;\!{\rm sin}[\eta({\bf p})]v_{0}(-{\bf p},h,\tau)(-\kappa)^{1/2},
\end{eqnarray}
where $v_{0}({\bf p},h,\tau)\equiv-i\gamma^{2}u_{0}^{\ast}({\bf p},h,\tau)$

\subsection{The preferred $CPT$ invariant vacuum}
\label{CPT_vacuum_subsec_spinor}       

In this subsection we show that, among the $CPT$-invariant vacua $|0_{\eta}\rangle$ on a time-reversal-symmetric FRW background, one particular vacuum is preferred: the $\eta=0$ vacuum $|0_{0}\rangle$.  This is again similar to the situation in de Sitter space where, among the de Sitter invariant vacua (the $\alpha$ vacua), one is preferred (the "Bunch-Davies" vacuum).

Here are two simple arguments that both lead us to the preferred vacuum $|0_{0}\rangle$:  
\begin{itemize}
\item First, consider $\langle 0_{\eta}|a_{\pm}^{\dagger}({\bf p},h)a_{\pm}^{}({\bf p},h)|0_{\eta}\rangle$ -- {\it i.e.}\ the expectation value of the number operator for particles of momentum ${\bf p}$ and helicity $h$, according to an asymptotic observer (long before or long after the bang), assuming that the universe is in the $CPT$-invariant vacuum state $|0_{\eta}\rangle$.  Using Eqs.~(\ref{Bogoliubov_pm_spinor}, \ref{B_pm_explicit_spinor}), we find
\begin{equation}
    \langle 0_{\eta}|a_{\pm}^{\dagger}({\bf p},\!h)a_{\pm}^{}({\bf p},\!h)|0_{\eta}\rangle
    \!=\!(2\pi)^{3}\delta({\bf 0})|\hat{\beta}_{\pm}(p)|^{2}\!\!\!\!\!\!\!\!\!\!
\end{equation}
where
\begin{equation}
  |\hat{\beta}_{\pm}(p)|^{2}=\frac{1-{\rm cos}[2\eta({\bf p})]{\rm cos}[\lambda({\bf p})]}{2}.
\end{equation}
Dividing out the uninteresting divergence $\delta({\bf 0})$ coming from the infinite spatial volume, and recalling from Subsection
(\ref{spinor_bogoliubov_section}) that $-\pi/2<\lambda({\bf p})<\pi/2$, we see that on a given FRW background spacetime ({\it i.e.}\ for fixed $\lambda$), the number density of particles (according to an asymptotic observer) is minimized when $\eta({\bf p})$ is an integral
multiple of $\pi$ or, without loss of generality, when $\eta({\bf p})=0$.
\item Second, consider the quantity $\langle 0_{\eta}|H|0_{\eta}\rangle$ -- {\it i.e.}\ the expectation value of the Hamiltonian 
$H$, according to an asymptotic observer, assuming the universe is in the $CPT$-invariant vacuum state $|0_{\eta}\rangle$.
From (\ref{Dirac_L}) we first obtain the Hamiltonian
\begin{equation}
  \label{H_psi}
  H=\int d^{3}x\Big[-i\bar{\psi}\gamma^{j}\partial_{j}\psi+\mu\bar{\psi}\psi\Big]
\end{equation}
where $j$ is summed from $1$ to $3$.  Then, using Eqs.~(\ref{psi_expansion_2}, \ref{spinor_bc}, \ref{u_from_upm_spinor}, \ref{B_pm_explicit_spinor}), we find that asymptotically (as $\tau\to\pm\infty$) $\langle 0_{\eta}|H|0_{\eta}\rangle$ is given by
\begin{equation}
  -\delta({\bf 0})\sum_{h}\int d^{\;\!3}p\;\omega(p,\tau){\rm cos}[2\eta({\bf p})]{\rm cos}[\lambda({\bf p})]\,.
\end{equation}
Once again, after dividing out $\delta({\bf 0})$,  and recalling that $-\pi/2<\lambda({\bf p})<\pi/2$, we see that on a given FRW background spacetime ({\it i.e.}\ for fixed $\lambda$), the energy density (according to an asymptotic observer) is minimized when $\eta({\bf p})$ is an integral multiple of $\pi$ or, without loss of generality, when $\eta({\bf p})=0$.
\end{itemize}
Thus, just as in the scalar case, we find that among the $CPT$-invariant vacua $|0_{\eta}\rangle$, the vacuum $|0_{0}\rangle$ is the one that is least excited (in the sense of having minimum expected particle density and minimum energy density).  


\section{The standard model and right-handed neutrinos}

Now that we have identified the $CPT$-invariant vacuum on a time-reversal-invariant FRW background, let us hypothesize that our universe is actually in this state, and work out some of the implications.  
We begin, in this section, 
by reviewing the standard model of particle physics, augmented by right-handed neutrinos, and their role as the only dark matter candidates in the theory.
Then, in the next section, we derive some pre- and postdictions for phenomenology and cosmology that follow from our perspective.

\subsection{The action}

\subsubsection{The standard model of particle physics, on curved spacetime}

We consider the standard model of particle physics (on an arbitrary curved spacetime background), with the usual gauge group $SU(3)_{C}\times SU(2)_{L}\times U(1)_{Y}$ and the usual matter fields (including a right-handed neutrino in each generation).  In this first subsection, we summarize this model for definiteness and to establish notation.  For a more thorough and pedagogical introduction to the standard model in flat spacetime, see {\it e.g.} Ref.~\cite{Langacker} (particularly Section 8.1); and for an explanation of how the coupling to gravity works (especially for spinor fields), see {\it e.g.} Section 3.8 in Ref.~\cite{BirrellDavies} or Section 12.5 in Ref.~\cite{WeinbergGR}.

The matter fields, and their $SU(3)\times SU(2)\times U(1)$ charges, are summarized in the following table:
\begin{equation}
  \label{SM_table}
  \begin{array}{c|c|c|c} 
    & SU(3)_{C} & SU(2)_{L} & U(1)_{Y} \\
    \hline
    q_{L}^{i} & 3 & 2 & +1/6 \\
    \hline
    u_{R}^{i} & 3 & 1 & +2/3 \\
    \hline
    d_{R}^{i} & 3 & 1 & -1/3 \\
    \hline
    l_{L}^{i} & 1 & 2 & -1/2 \\
    \hline
    \nu_{R}^{i} & 1 & 1 & 0 \\
    \hline
    e_{R}^{i} & 1 & 1 & -1 \\
    \hline
    h & 1 & 2 & +1/2 \\
  \end{array}
\end{equation}
Here $q_{L}$ is the left-handed quark doublet, $u_{R}$ and $d_{R}$ are the corresponding right-handed quarks; 
$l_{L}$ is the left-handed lepton doublet, $\nu_{R}$ and $e_{R}$ are the corresponding right-handed leptons; and $h$
is the Higgs doublet.  The superscript "$i$" on the spinor fields runs from 1 to 3, and is a reminder that the standard model fermions come in three families.

We take the most general renormalizable action for these fields on a curved background, $S=\int d^{4}x\sqrt{-g} L$, with Lagrangian:
\begin{eqnarray}
  \label{SM_action}
    L\!\!&\!=\!&\!\!\frac{R\!-\!2\Lambda}{16\pi G}\!+\!\xi R h^{\dagger}h
    \!-\!(D_{\mu}h)^{\dagger}(D^{\mu}h)\!-\!m^{2}h^{\dagger}h\!-\!\frac{\lambda_{h}}{4}(h^{\dagger}h)^{2} 
    \nonumber\\
    \!\!&\!-\!&\!\!\frac{1}{4}{\rm Tr}[G_{\mu\nu}G^{\mu\nu}]-\frac{1}{4}{\rm Tr}[W_{\mu\nu}W^{\mu\nu}]-\frac{1}{4}B_{\mu\nu}B^{\mu\nu} 
    \nonumber\\
    \!\!&\!+\!&\!\!i\left(
    \bar{q}_{L}D\!\!\!\!/\,q_{L}+\bar{u}_{R}D\!\!\!\!/\,u_{R}+\bar{d}_{R}D\!\!\!\!/\,d_{R}\right.
    \nonumber\\
    \!\!&\!\!&\!\!\;\left.+\bar{l}_{L}D\!\!\!\!/\,l_{L}\;\!+\bar{\nu}_{R}D\!\!\!\!/\,\nu_{R}\;\!+\bar{e}_{R}D\!\!\!\!/\,e_{R}\right) 
    \nonumber\\
    \!\!&\!-\!&\!\!(\bar{q}_{L}Y_{u}^{\dagger}u_{R}\tilde{h}
   \!+\!\bar{q}_{L}Y_{d}^{\dagger}d_{R}h
   \!+\!\bar{l}_{L}Y_{\nu}^{\dagger}\nu_{R}\tilde{h}
   \!+\!\bar{l}_{L}Y_{e}^{\dagger}e_{R}h\!+\!{\rm h.c.})
   \nonumber\\
   \!\!&\!-\!&\!\!\frac{1}{2}(\bar{\nu}_{R}^{c}M^{\dagger}\nu_{R}+{\rm h.c.})\Big\}.
\end{eqnarray}
Here $R$ is the Ricci scalar; $\Lambda$ is the cosmological constant; $G$ is Newton's constant; $\{\xi,m,\lambda_{h}\}$ are real-valued constants; $D_{\mu}$ is the gauge-covariant derivative on $h$; $G_{\mu\nu}$, $W_{\mu\nu}$ and $B_{\mu\nu}$ are the $SU(3)$, $SU(2)$ and $U(1)$ field strength tensors; $D\!\!\!\!/$ is the covariant Dirac operator (including both the gauge and gravitational connection); $\tilde{h}=i\sigma^{2}h^{\ast}$, where $\sigma^{2}$ is the second Pauli sigma matrix; $Y_{u}$, $Y_{d}$, $Y_{\nu}$, $Y_{e}$ and $M$ are constant $3\times3$ complex matrices, contracted with the family indices on the fermion fields; and $\nu_{R}^{c}=-i\gamma^{2}\nu_{R}^{\ast}$ is the charge conjugate of the $\nu_{R}$.  So: {\it (i)} the first line has the gravitational terms and the kinetic and potential terms for the Higgs; {\it (ii)} the second line has the gauge kinetic terms (there are also the parity violating gauge kinetic terms which we have suppressed); {\it (iii)} the third line has the kinetic terms for the fermions; {\it (iv)} the fourth line has the Yukawa couplings; and {\it (v)} the fifth line has the majorana mass term for the right-handed neutrinos. 

Note that we have omitted terms that are incompatible with second-order equations of motion (in particular, the Weyl-curvature-squared term), and suppressed the topological terms (the Gauss-Bonnet term, and the $G\tilde{G}$ dual terms).  Also note that on the first line of the action, in addition to the standard Einstein-Hilbert term and the cosmological constant, we have included the term $\xi R h^{\dagger}h$: there is no experimental or theoretical reason to exclude this term.  In fact, on the theoretical side, if we want to treat the standard model as a QFT on a classical curved background spacetime (which is the simplification we take in this paper), we must include this term in order for the theory to be renormalizable \cite{Callan:1970ze}; and on the experimental side, the constraints on $\xi$ are very weak ($|\xi|<2.6\times10^{15}$ \cite{Atkins:2012yn}).  

This action is $CPT$ invariant on Minkowski space or on a flat FRW background with time-reversal symmetry.

\subsubsection{Right-handed neutrinos, $\mathbb{Z}_{2}$ symmetry, and dark matter}

We emphasize that we have included a right-handed neutrino in each generation.  The three right-handed neutrinos have not yet been observed, but there are several reasons to expect that they exist: 
\begin{enumerate}
\item They provide the simplest renormalizable explanation for the observed neutrino oscillations and for the smallness of the light neutrino masses (via the see-saw mechanism). 
\item They offer a natural explanation for the observed matter/anti-matter asymmetry: thermal leptogenesis \cite{Fukugita:1986hr, Buchmuller:2005eh}; and this suggests a heavy neutrino mass scale that agrees with the one suggested by the see-saw mechanism (the GUT scale, roughly).  
\item They complete the natural pattern of particles in the standard model (so that every left-handed particle has a right-handed partner, and the lepton representations $\{l_{L},\nu_{R},e_{R}\}$ are a colorless analogue of the quark representations
$\{q_{L},u_{R},d_{R}\}$).
\item When we include the right-handed neutrinos, the 16 Weyl fermions in each standard model generation, with their assorted $SU(3)\times SU(2)\times U(1)$ charge assignments, naturally unify into a single irrep: the Weyl spinor of $SO(10)$.  
\item Finally, as we shall see, in our $CPT$-invariant picture, one of the right-handed neutrinos also becomes a good dark matter candidate.
\end{enumerate}

Indeed, looking through the various particles in the above theory (\ref{SM_action}), we see that there is only one possible dark matter candidate -- {\it i.e.}\ only one particle that, on the one hand, has not yet been detected and, on the other hand, can have a lifetime longer than the Hubble time.  In particular, current experimental constraints allow for the possibility that one of the three right-handed neutrinos, $\nu_{R}$, is exactly stable.  (Note that {\it at most} one of the heavy right-handed neutrinos can be stable since, for every heavy right-handed neutrino that is stable, there is a corresponding light left-handed neutrino that is massless, and we know observationally that at most one of the light neutrinos is massless.)  

We choose a flavor basis where the symmetric $3\times3$ "majorana" mass matrix $M$ is diagonal, with eigenvalues $\{M_{1},M_{2},M_{3}\}$; and where the "first" right-handed neutrino $\nu_{R,1}^{}$ (with eigenvalue $M_{1}$) is the stable one.  In order for $\nu_{R,1}^{}$ to be stable, the first row of the matrix $Y_{\nu}$ must vanish (since a non-zero value for the $j$th element in this row would lead to an unwanted decay $\nu_{R,1}^{}\to l_{L,j}+h$).  If the first row of $Y_{\nu}$ vanishes, the action (\ref{SM_action}) automatically has an extra $\mathbb{Z}_{2}$ symmetry under $\nu_{R,1}^{}\to-\nu_{R,1}^{}$.  Stated another way, we can set the first row of $Y_{\nu}$ to zero (and hence stabilize the first right-handed neutrino $\nu_{R,1}^{}$) by demanding that the action (\ref{SM_action}) has a global $\mathbb{Z}_{2}$ symmetry under $\nu_{R,1}^{}\to-\nu_{R,1}^{}$. 

The $\nu_{R,1}\to-\nu_{R,1}$ symmetry implies that the stable neutrino $\nu_{R,1}$ only interacts with gravity: its interactions with the other standard model fields vanish, and it may be thought of as a free fermion living on the spacetime background.  At first glance, this would not seem like a viable dark matter candidate, since it is not produced by the thermal bath in the early universe.  But, as we will show, in our scenario it is instead produced by gravity (as a result of the inequivalence between the $CPT$-invariant state of the universe, and the "out" vacuum used by a late-time observer).

The idea that the dark matter is a right-handed (sterile) neutrino has a long history.  In its earlier incarnation \cite{Dodelson:1993je, Shi:1998km, Dolgov:2000ew, Abazajian:2001nj, Asaka:2005an}, the idea was that the dark matter neutrino had a non-trivial mixing with the active/light neutrinos, so that that it was produced by the thermal bath in the early universe, and was slightly unstable (and hence could decay at a rate that was potentially observable by x-ray telescopes).  It was necessarily a warm dark matter candidate with a mass in the $1-100~{\rm keV}$ range.  The simplest incarnation of this idea (based on non-resonant product, with a standard thermal history and negligible initial lepton asymmetry) is now ruled out by observational constraints, but a more involved scenario assuming resonant production and a significant initial lepton asymmetry (much larger than the observed baryon asymmetry) may still be viable \cite{Abazajian:2006yn, Boyarsky:2009ix, Canetti:2012kh}. In any case, we emphasize that, although our dark matter candidate is also a sterile neutrino, our scenario is otherwise completely different from this earlier one: our dark matter candidate is  much colder; is much heavier ($4.8\times10^{8}~{\rm GeV}$ as opposed to $\sim 10~{\rm keV}$); can be (and likely is) completely stable and completely decoupled from the rest of the standard model; and is produced by gravity rather than by the thermal radiation bath.

\subsection{Weyl-invariant reformulation} 
\label{Weyl}

In this subsection, we will reformulate the action (\ref{SM_action}) in the equivalent but more symmetric form (\ref{SM_action_Weyl}): the reformulated action contains an extra scalar field (the dilaton field $\varphi$), as well as an extra gauge invariance (under local scale transformations, or "Weyl transformations"), so that the total number of degrees of freedom is unchanged.  From the physical standpoint, this reformulation implements the idea (originally emphasized by Weyl and then Dicke) that, just as physics should not depend on the local coordinate frame used to describe it, it should not depend on the local choice of units used to describe it.  From the mathematical standpoint, this reformulation is just the gravitational version of the familiar Stueckelberg trick \cite{Stueckelberg:1900zz, Stueckelberg:1938zz, Codello:2012sn}.  Let us begin by explaining these two perspectives in a bit more depth:

\subsubsection{Physical perspective}  A key principle underlying general relativity (GR) is the idea that spacetime coordinates are mere labels: two different coordinate systems just correspond to two different conventions about how to describe the same underlying physical configuration, and physics may be formulated in a "covariant" way that makes this fact manifest.  Moreover, this coordinate freedom is {\it local}: choosing the coordinate frame near a spacetime point $p$ does not fix the coordinate frame near a different spacetime point $p'$.  

Soon after Einstein introduced GR, Weyl \cite{Weyl:1918ib} (and later Dicke \cite{Dicke:1961gz}) observed that the same line of thinking ought to extend to the choice of the units as well: two different choices for the unit of length are just two different conventions about how to describe the same physical configuration, and the laws of physics should be formulated in a way that makes this fact manifest, even when the choice of unit length is regarded as a local one.  But note that, as it is {\it traditionally} formulated, Einstein's theory does {\it not} include the freedom to choose units locally: as soon as Alice, at one spacetime point $p$, picks up a rod and declares "This is one meter!", the notion of "one meter" is suddenly defined globally; and Bob, at a different spacetime point $p'$, with several rods of different lengths in front of him, is no longer is free to choose which of them is one meter long.  After all, in the traditional formulation, the electron is taken to have the same mass or, equivalently, the same Compton wavelength, $\approx 2.426\times10^{-12}{\rm m}$, at every point in spacetime.  

Both Weyl and Dicke argue that this lack of local freedom in the choice of units is a deficiency in the traditional formulation of GR: it does not seem to have an empirical basis, and it is at odds with the idea of "infinitessimal geometry" that is so central to GR.  As Weyl observes: in GR, we cannot compare vectors a distant spacetime points, so why should we be able to compare their length?  As Dicke puts it: "Imagine, if you will, that you are told by a space traveller that a hydrogen atom on Sirius has the same diameter as one on the earth.  A few moments' thought will convince you that the statement is either a definition or else meaningless.  It is evident that two rods side by side, stationary with respect to each other, can be intercompared and equality established in the sense of an approximate congruence between them.  However, this cannot be done for perpendicular rods, for rods moving relatively, or for rods with either a space- or time-like separation."

\subsubsection{Mathematical perspective} 

Now we explain how to promote any action $S$ to the corresponding equivalent Weyl-invariant action $\tilde{S}$ in which the local freedom to change units is manifest.  This is achieved by the gravitational version of the Stueckelberg trick.\footnote{In Weyl's original paper \cite{Weyl:1918ib}, he presented a general technique for promoting a theory $S$ that is diff invariant to a corresponding theory $\tilde{S}$ that is both diff and Weyl invariant, by gauging the group of scale transformations.  In Weyl's approach, the new (gauged) theory $\tilde{S}$ contains an extra gauge field $b_{\mu}$, and thus is not equivalent to the original (ungauged) theory.  By contrast, the Stueckelberg trick described in this subsection produces a new theory $\tilde{S}$ that {\it is} equivalent to the original theory $S$.  It was later understood that the Stueckelberg trick corresponds to the special case of Weyl's technique in which the new gauge boson is pure gauge: $b_{\mu}=\partial_{\mu}\chi$.  See Ref.~\cite{Codello:2012sn} for a nice explanation of this point, and various other aspects of the issues issues discussed in this subsection and the previous one.}

First recall the original Stueckelberg trick.  Consider the action for a massive vector field $A_{\mu}$:
\begin{equation}
  S=\int d^{4}x\left[-\frac{1}{4}F_{\mu\nu}F^{\mu\nu}-\frac{1}{2}m^{2}A_{\mu}A^{\mu}\right]
\end{equation}  
where $F_{\mu\nu}=\partial_{\mu}A_{\nu}-\partial_{\nu}A_{\mu}$.  The ordinary gauge invariance under $A_{\mu}\to A_{\mu}+\partial_{\mu}\chi$ is spoiled here by the presence of the mass term $-m^{2}A_{\mu}A^{\mu}$.  To restore this gauge symmetry, we introduce a new "Stueckelberg" field $\varphi$ via the substitution $A_{\mu}\to A_{\mu}+\partial_{\mu}\varphi$ to obtain the new action
\begin{equation}
  \tilde{S}=\int d^{4}x\left[-\frac{1}{4}F_{\mu\nu}F^{\mu\nu}-\frac{1}{2}m^{2}(A_{\mu}+\partial_{\mu}\varphi)(A^{\mu}+\partial^{\mu}\varphi)\right].
\end{equation}
This new action $\tilde{S}$ is invariant under the simultaneous gauge transformation $A_{\mu}\to A_{\mu}+\partial_{\mu}\chi$ and $\varphi\to\varphi-\chi$; and it is equivalent to the original action $S$ (which is recovered in "unitary gauge" $\varphi={\rm constant}$).

Now we turn to the gravitational version.  Consider an arbitrary diff-invariant action $S[\Theta_{a}]$ that is a function of various fields $\Theta_{a}$.  We take $\hbar=c=1$ so that every field $\Theta_{a}$ has dimensions of length to some power $w_{a}$; and we take the spacetime coordinates to be dimensionless labels, so that $g_{\mu\nu}$ has dimensions of length squared.  Now imagine we shrink our chosen unit of length by a spacetime-dependent factor $\Omega(x)$, so each field gets rescaled by an appropriate power of $\Omega$, depending on its length dimension: $\Theta_{a}\to\Omega^{w_{a}}\Theta_{a}$.\footnote{In particular, we have $g_{\mu\nu}\to\Omega^{2}g_{\mu\nu}$ (for the metric), $e_{\mu}^{a}\to\Omega^{1}e_{\mu}^{a}$ (for the veirbein fields), $\phi\to\Omega^{-1}\phi$ (for a scalar field), $\psi\to \Omega^{-3/2}\psi$ (for a fermion field), and $A_{\mu}\to\Omega^{0}A_{\mu}$ (for a gauge field).}  Now, we say an action $S$ is "Weyl invariant" if it is invariant under such a rescaling: $S[\Theta_{a}]=S[\Omega^{w_{a}}\Theta_{a}]$.  If $S$ is {\it not} Weyl invariant, we again introduce a Stueckelberg field $\varphi$ via a substitution modeled on the desired local transformation law: $\Theta_{a}\to(\varphi/\hat{\mu})^{w_{a}}\Theta_{a}$, where in this case the Stueckelberg field $\varphi$ is known as the dilaton, and $\hat{\mu}$ is an arbitrary mass scale that may be chosen for convenience and is inserted so that $\varphi$ has dimensions of ${\rm mass}={\rm length}^{-1}$.  In this way, we obtain a new action $\tilde{S}[\Theta_{a},\varphi]=S[(\varphi/\mu)^{w_{a}}\Theta_{a}]$ that: (i) is Weyl invariant (under the Weyl transformation $\Theta_{a}\to\Omega^{w_{a}}\Theta_{a}$ and $\varphi\to\Omega^{-1}\varphi$); and (ii) is equivalent to the original action $S$ (which is recovered in "unitary gauge" $\varphi=\hat{\mu}$). 

If we apply this technique to our action (\ref{SM_action}), we obtain the new Weyl-invariant action $\tilde{S}=\int d^{4}x \tilde{L}$, with Lagrangian
\begin{eqnarray}
  \label{SM_action_Weyl}
    \tilde{L}
    \!&\!=\!&\!\frac{1}{2}\!\Big[(\partial\varphi)^{2}\!+\!\frac{1}{6}\varphi^{2}R\Big]\!
    +6\xi\!\Big[|Dh|^{2}\!+\!\frac{1}{6}|h|^{2}R\Big]
    \nonumber \\
    \!&\!-\!&\!(6\xi\!+\!1)\Big|Dh\!-\!(h/\varphi)\partial\varphi\Big|^{2}
    \nonumber \\
    \!&\!-\!&\!(1/4)\lambda_{h}|h|^{4}-(1/2)\lambda_{m}|h|^{2}\varphi^{2}-(1/4)\lambda_{\varphi}\varphi^{4} 
    \nonumber \\
    \!&\!-\!&\!\frac{1}{4}{\rm Tr}[G_{\mu\nu}G^{\mu\nu}]-\frac{1}{4}{\rm Tr}[W_{\mu\nu}W^{\mu\nu}]-\frac{1}{4}B_{\mu\nu}B^{\mu\nu} 
    \nonumber\\
    \!&\!+\!&\!i\left(
    \bar{q}_{L}D\!\!\!\!/\,q_{L}
    +\bar{u}_{R}D\!\!\!\!/\,u_{R}
    +\bar{d}_{R}D\!\!\!\!/\,d_{R}\right. \nonumber\\
    \!&\!+\!&\!\;\;\;\;\left.\bar{l}_{L}D\!\!\!\!/\,l_{L}
    \;\!+\bar{\nu}_{R}D\!\!\!\!/\,\nu_{R}
    \;\!+\bar{e}_{R}D\!\!\!\!/\,e_{R}\right) 
    \nonumber\\
    \!&\!-\!&\!(\bar{q}_{L}Y_{u}^{\dagger}u_{R}\tilde{h}
   \!+\!\bar{q}_{L}Y_{d}^{\dagger}d_{R}h
   \!+\!\bar{l}_{L}Y_{\nu}^{\dagger}\nu_{R}\tilde{h}
   \!+\!\bar{l}_{L}Y_{e}^{\dagger}e_{R}h\!+\!{\rm h.c.})
   \nonumber\\
   \!&\!-\!&\!\frac{1}{2}\varphi(\bar{\nu}_{R}^{c}Y_{m}^{\dagger}\nu_{R}+{\rm h.c.}).
\end{eqnarray}
Here, without loss of generality, we have chosen the arbitrary mass scale $\hat{\mu}$ to be 
\begin{equation}
\label{mu_hat}
  \hat{\mu}=(4\pi G/3)^{-1/2} \approx 5.966\times10^{18}~{\rm GeV}, 
\end{equation}
so that the kinetic term for $\varphi$ is canonically normalized in the special case ($\xi=-1/6$) where the $\varphi$ and $h$ kinetic terms decouple from one another.  Note that, in the new Weyl-invariant action (\ref{SM_action_Weyl}), all of the coupling constants are dimensionless.  The new dimensionless couplings are related to the previous dimensionful ones as follows: 
\begin{equation}
  \lambda_{m}=2 m^{2}/\hat{\mu}^{2},\quad 
  \lambda_{\varphi}=(2/3)\Lambda/\hat{\mu}^{2},\quad
  Y_{m}=M/\hat{\mu}.
\end{equation}
In particular, since $M={\rm diag}\{M_{1},M_{2},M_{3}\}$, $Y_{m}={\rm diag}\{y_{m,1},y_{m,2},y_{m,3}\}$ where 
\begin{equation}
\label{y_m_i}
y_{m,i}=M_{i}/\hat{\mu}.
\end{equation}

We emphasize that the freedom to locally choose the unit of length also extends to the quantum theory (where it is again implemented by the same Stuckelberg trick).  The corresponding quantization procedure and associated RG flow preserve Weyl-invariance (see Ref.~\cite{Codello:2012sn} and references therein).  In particular, one can proceed from the action $S[\Theta_{a}]$ to the corresponding Weyl-invariant quantum effective action $\tilde{\Gamma}[\Theta_{a},\varphi]$ via either of the following two (equivalent) paths:  
\begin{itemize}
\item Path 1: $S[\Theta_{a}]\to\tilde{S}[\Theta_{a},\varphi]\to\tilde{\Gamma}[\Theta_{a},\varphi]$.  In other words, first apply the Stueckelberg trick to $S[\Theta_{a}]$ to obtain the equivalent Weyl-invariant action $\tilde{S}[\Theta_{a},\varphi]$, and then quantize $\tilde{S}[\Theta_{a},\varphi]$ to obtain the corresponding quantum effective action $\tilde{\Gamma}[\Theta_{a},\varphi]$;
\item Path 2: $S[\Theta_{a}]\to\Gamma[\Theta_{a}]\to\tilde{\Gamma}[\Theta_{a},\varphi]$.  In other words, first quantize $S[\Theta_{a}]$ to obtain the corresponding quantum effective action $\Gamma[\Theta_{a}]$, and then apply the Stueckelberg trick to $\Gamma[\Theta_{a}]$ to obtain the equivalent Weyl-invariant quantum effective action $\tilde{\Gamma}[\Theta_{a},\varphi]$.
\end{itemize}
These two paths yield the same result for $\tilde{\Gamma}[\Theta_{a},\varphi]$: in other words, quantization commutes with the Stueckelberg trick.  Again, see Ref.~\cite{Codello:2012sn} for a derivation of these results, and a clear explanation of how they are perfectly compatible with well-known results about Weyl anomalies and running couplings.

\section{Predictions for cosmology and particle physics} 
 
\subsection{Neutrino dark matter from CPT}
\label{Neutrino_DM}

In the previous subsection, we introduced the action (\ref{SM_action_Weyl}) for the standard model of particle physics on a curved spacetime background (written in an equivalent Weyl-invariant form).

As explained above, if the action is invariant under the $\mathbb{Z}_{2}$ symmetry $\nu_{R,1}\to-\nu_{R,1}$, the first right-handed neutrino $\nu_{R,1}$ becomes stable.  Let us study how this stable neutrino $\nu_{R,1}^{}$ behaves on a cosmological (spatially-flat FRW) background.  The flat FRW metric $g_{\mu\nu}=a^{2}\eta_{\mu\nu}$ is proportional to the Minkowski metric $\eta_{\mu\nu}$; the scale factor $a$ and the background scalar fields $h$ and $\varphi$ are purely functions of the conformal time $\tau$: $a=a(\tau)$, $h=h(\tau)$, $\varphi=\varphi(\tau)$; and the relevant Weyl transformations are those that preserve FRW: $\Omega=\Omega(\tau)$.  

Under a Weyl transformation $g_{\mu\nu}\to\Omega^{2}g_{\mu\nu}$, the relevant quantities transform as follows: $a\to\Omega^{+1}a$, $h\to\Omega^{-1}h$, $\varphi\to\Omega^{-1}\varphi$ and $\nu_{R,1}^{}\to\Omega^{-3/2}\nu_{R,1}^{}$.  It is therefore helpful to switch to the Weyl invariant combinations:
\begin{equation}
  H=ah,\qquad \Phi=a\varphi,\qquad N_{R,1}^{}=a^{3/2}\nu_{R,1}^{}.
\end{equation}

In this section, we proceed in two steps.  First, in Subsection \ref{neutrino_bogoliubov}, we follow the sterile (dark matter) neutrino's evolution through the bang, and determine the Bogoliubov transformation between the ingoing (pre-bang) and outgoing (post-bang) creation and annihilation operators.  Then, in Subsection \ref{dark_matter}, we discuss the resulting predictions for the dark matter' mass, adiabaticity, and coldness.

\subsubsection{Neutrino evolution through the bang}
\label{neutrino_bogoliubov}

Let us first consider the background evolution (near the bang, above the electroweak phase transition).  If we use the spatially-flat FRW expression $R=6 a''/a^{3}$ the action (\ref{SM_action_Weyl}) becomes
\begin{eqnarray}
  S\!\!&\!=\!&\!\!V_{3}\!\!\int\!\!d\tau\Big\{\!\!-\!\frac{1}{N}\!\Big[\frac{1}{2}\Phi'{}^{2}\!\!+\!6\xi\big|H'\big|^{2}\!\!\!-\!(6\xi\!+\!1)\!\big|H'\!-\!(H/\Phi)\Phi'\big|^{2}\Big]
  \nonumber\\
  \!\!&\!\!&\!\!\qquad\quad\;\;-N\Big[V(H,\Phi)+a^{4}\rho\Big]\Big\}
\end{eqnarray}
where $V_{3}$ is the comoving spatial volume, $N$ is the lapse function, the potential $V(H,\Phi)$ is
\begin{equation}
  V(H,\Phi)=\frac{1}{4}\lambda_{h}|H|^{4}+\frac{1}{2}\lambda_{m}|H|^{2}\Phi^{2}+\frac{1}{4}\lambda_{\varphi}\Phi^{4},
\end{equation}
and the remaining (gauge and fermion field) terms are swept into the effective energy density $a^{4}\rho$ term.  Note that during
the radiation era, $\rho\propto a^{-4}$ so that $a^{4}\rho$ is just a constant $\rho_{1}$.  We can vary the lapse function $N$ to obtain 
\begin{equation}
\label{Friedmann}
  \frac{1}{2}\Phi'{}^{2}+6\xi\big|H'\big|^{2}-(6\xi+1)\big|H'-(H/\Phi)\Phi'\big|^{2}=V(H,\Phi)+\rho_{1}.
\end{equation}
This is just the Friedmann equation written in less familiar variables.  

Now let us focus on the regime near the bang (above the electroweak phase transition).  In this regime, the background value of $H$ is pinned at zero by thermal effects, so we can set $H=H'=0$ in Eq.~(\ref{Friedmann}), and regard any thermal fluctuations in $H$ as incorporated in the radiation density $\rho$.  The $\lambda_{\varphi}\Phi^{4}$ term is just the usual (tiny) cosmological constant term in the Friedmann equation, written in an unfamiliar form: during the radiation era it is utterly negligible compared to the radiation density, and may be ignored.  Thus, above the electroweak phase transition, Eq.~(\ref{Friedmann}) simply reduces to $(1/2)\Phi'{}^{2}=\rho_{1}$, so the solution is
\begin{equation}
  \label{Phi(tau)}
  \Phi(\tau)=(2\rho_{1})^{1/2}\tau.
\end{equation}
(This is the familiar statement that $a(\tau)\propto\tau$ during the radiation-dominated era.)  

Now let us consider the evolution of the stable dark-matter neutrino species on this background.  From Eq.~(\ref{SM_action_Weyl}), we find that $N_{R,1}^{}$ obeys the Majorana-like equation of motion
\begin{equation}
  \label{Majorana_N}
  i\partial\!\!\;\!\!\!/\,N_{R,1}^{}=y_{m,1}^{}\Phi N_{R,1}^{c}
\end{equation}
where $\partial\!\!\!\;\!\!/\,=\gamma^{\mu}\partial_{\mu}$ is the ordinary flat-space Dirac operator, $y_{m,1}$ is given by Eq.~(\ref{y_m_i}), and $N_{R,1}^{c}\equiv-i\gamma^{2}N_{R,1}^{\ast}$.  It is convenient to rewrite Eq.~(\ref{Majorana_N}) in the equivalent Dirac-like form
\begin{equation}
  \label{Dirac_N}
  (i\partial\!\!\;\!\!\!/\,-\mu)N_{1}=0,
\end{equation}
where we have defined the Majorana spinor
\begin{equation}
  \label{mu}
  N_{1}^{}\equiv N_{R,1}^{}+N_{R,1}^{c}
\end{equation}
and its effective mass $\mu(\tau)\equiv y_{m,1}^{}\Phi(\tau)$.  With the solution (\ref{Phi(tau)}) for $\Phi(\tau)$, we see that the effective neutrino mass $\mu(\tau)$ is given by
\begin{equation}
  \label{mu_spinor}
  \mu(\tau)=\gamma\tau
\end{equation}
with constant coefficient
\begin{equation}
  \label{gamma}
  \gamma\equiv y_{m,1}^{}(2\rho_{1}^{})^{1/2}.
\end{equation}

Our goal is to solve the Dirac equation (\ref{Dirac_N}).  We can proceed by finding all solutions, and then restricting to the Majorana solutions (satisfying $N_{1}=-i\gamma^{2}N_{1}^{\ast}$) at the end.

Now, before analyzing the exact solution of (\ref{Dirac_N}), let us begin by thinking about how the solution should behave in the large $p$ and small $p$ limits:
\begin{itemize}
\item Large $p$ limit.  The comoving frequency is: $\omega^{2}=p^{2}+\mu^{2}=p^{2}+(2\rho_{1})y_{m,1}^{2}\tau^{2}$.  Thus, for fixed $p$, the dimensionless WKB parameter $\omega'/\omega^{2}$ reaches a maximal value 
\begin{subequations}
  \label{wkb}
  \begin{equation}
    \left.\frac{\omega'}{\omega^{2}}\right|_{\rm max}=\left(\frac{2}{3}\right)^{3/2}\frac{y_{m,1}\rho_{1}^{1/2}}{p^{2}}
  \end{equation}
  at a time
  \begin{equation}
    \tau_{p}=\frac{p}{2\;\!y_{m,1}\rho_{1}^{1/2}}.
  \end{equation}
\end{subequations}
From here we see that, in the Bogoliubov transformation between the ingoing and outgoing modes, the Bogoliubov $\beta$ coefficient should vanish exponentially for $p^{2}\gg\gamma$ (since the WKB parameter is always much less than unity for such large $p$ modes).
\item Small $p$ limit.  In the $p\to0$ limit, Eq.~(\ref{Dirac_N}) becomes $[i\gamma^{0}\partial_{\tau}-\mu(\tau)]N_{1}=0$, which has the general exact solution
\begin{eqnarray}
  N_{1}(\tau)&=&{\rm exp}\left[+i\int^{\tau}\mu(\tilde{\tau})d\tilde{\tau}\right]\left(\begin{array}{c} \,\,\xi_{+} \\ \!\!+\xi_{+} \end{array}\right) \nonumber\\
  &+&{\rm exp}\left[-i\int^{\tau}\mu(\tilde{\tau})d\tilde{\tau}\right]\left(\begin{array}{c} \,\,\xi_{-} \\ \!\!-\xi_{-} \end{array}\right),
\end{eqnarray}
where $\xi_{+}$ and $\xi_{-}$ are arbitrary 2-component constant spinors.  From this solution we see that, as we pass 
through the bang and $\mu$ switches from negative to positive, the solution proportional to ${\rm exp}[+i\int^{\tau}\mu(\tilde{\tau})d\tilde{\tau}]$ switches from positive frequency to negative frequency, while the solution proportional to ${\rm exp}[-i\int^{\tau}\mu(\tilde{\tau})d\tilde{\tau}]$
switches from negative frequency to positive frequency.  Thus, the Bogoliubov $\beta$ coefficient should be maximal 
($|\beta|\to1$) in the long-wavelength ($p\to0$) limit.
\end{itemize}
Now let us return to the exact solution of Eq.~(\ref{Dirac_N}).  In Appendix \ref{exact_bogoliubov}, we show that the exact solutions of Eq.~(\ref{Dirac_N}) may be expressed in terms of parabolic cylinder functions $D_{p}(z)$; and that, by using the known asymptotic expansions for these functions, we can determine the exact Bogoliubov transformation between the ingoing and outgoing positive and negative frequency solutions.  In particular, we find that that Bogoliubov $\beta$ coefficient between the ingoing and outgoing modes (which measures the inequivalence between the corresponding in and out vacua) satisfies:
\begin{equation}
  \label{beta_in_out}
  \left|\beta({\bf p},h)\right|={\rm exp}\left(-\frac{\pi p^{2}}{2\gamma}\right).
\end{equation}
As a check, we note that this exact formula has the limiting behavior expected from our approximate arguments above: $|\beta|$ approaches zero exponentially in the short-wavelength limit $p\gg\gamma^{1/2}$, and $|\beta|$ approaches unity in the long-wavelength limit $p\ll\gamma^{1/2}$.

\subsection{Dark matter predictions}
\label{dark_matter}

Now we discuss our dark matter predictions: we predict the mass of the dark matter particle; and we explain why our mechanism automatically predicts that the dark matter is adiabatic and cold, with ultra-weak interactions.

\subsubsection{Dark matter mass}  

Eq.~(\ref{beta_in_out}) gives the size of the Bogoliubov coefficient $|\beta({\bf p},h)|=|{\rm sin}[\lambda({\bf p})]|$ (between the ingoing vacuum and the outgoing vacuum).  Then, using the results of Subsection \ref{CPT_vacuum_subsec_spinor}, we infer the size of the corresponding Bogoliubov coefficient $|\beta_{+}({\bf p},h)|=|{\rm sin}[\lambda({\bf p})/2]|$ (between the CPT-invariant vacuum and the outgoing vacuum):
\begin{equation}
  \left|\beta_{+}({\bf p},h)\right|= {\rm sin}\Big\{\frac{1}{2}{\rm arcsin}\big[\left|\beta({\bf p},h)\right|\big]\Big\},
\end{equation}
so that the comoving number density of dark matter neutrinos (according to late-time observers like us, assuming the universe is {\it actually} in the CPT invariant state) is
\begin{equation}
  \label{n_dm}
  n_{dm}=\sum_{h=1,2}\int \frac{d^{3}{\bf p}}{(2\pi)^{3}}\left|\beta_{+}({\bf p},h)\right|^{2}=(\gamma/\pi)^{3/2}I
\end{equation} 
where $I$ is a dimensionless constant defined as follows:
\begin{equation}
  I\;\equiv\;\frac{1}{2\pi^{2}}\int_{0}^{\infty}dx\,x^{2}\left[1-\sqrt{1-{\rm e}^{-x^{2}}}\,\right]\;\approx\;0.01276.
\end{equation}
To compare with observations, we want a quantity that will not dilute over time, so we consider the dark matter yield
\begin{equation}
  Y_{dm}=n_{dm}/s,
\end{equation}
where now $n_{dm}$ and $s$ are the dark matter number density and total entropy density, respectively, at any moment in the cosmic expansion.  In the early universe (after the decay of the two {\it unstable} heavy neutrino species, but before the electroweak phase transition) the energy density and entropy density were dominated by the radiation fluid, with
\begin{subequations}  
  \begin{eqnarray}
    \rho&=&\frac{\pi^{2}}{30}g_{\ast}T^{4}, \\
    s&=&\frac{2\pi^{2}}{45}g_{\ast}T^{3},
  \end{eqnarray}
\end{subequations}
and hence
\begin{equation}
  \label{rho_over_s}
  \frac{(2\rho)^{3/4}}{s}=\frac{3}{2}\left(\frac{15}{g_{\ast}\pi^{2}}\right)^{1/4},
\end{equation}
where $g_{\ast}=106.75$ is the number of effective degrees of freedom in the standard model, excluding the right-handed neutrinos (see Ch.~3 in Ref.~\cite{KolbTurner}).  By combining Eqs.~(\ref{y_m_i}, \ref{gamma}, \ref{n_dm}, \ref{rho_over_s}), we obtain the following expression for the dark matter yield $Y_{dm}$:
\begin{equation}
  \label{Ydm}
  Y_{dm}=\frac{3I}{2\pi^{2}}\left(\frac{15}{g_{\ast}}\right)^{1/4}\left(\frac{M_{1}}{\hat{\mu}}\right)^{3/2}
\end{equation} 
where $M_{1}$ is the dark matter mass, and $\hat{\mu}$ is given by Eq.~(\ref{mu_hat}).  The predicted present-day dark matter energy density is then
\begin{equation}
  \label{rho_dm}
  \rho_{dm}^{(0)}=M_{1}^{}n_{dm}^{(0)}=M_{1}^{}Y_{dm}^{}s^{(0)}
\end{equation}
where $s^{(0)}\approx 2.3\times10^{-38} {\rm GeV}^{3}$ is the present entropy density \cite{KolbTurner}.  If we equate this prediction to the {\it observed} value of the present-day dark matter density \cite{Ade:2015xua}
\begin{equation}
  \rho_{dm}^{(0)}\approx 9.7\times10^{-48}~{\rm GeV}^{4},
\end{equation} 
we find that the dark matter neutrino must have mass:
\begin{equation}
  \label{M_1}
  M_{1}^{}=4.8\times10^{8}~{\rm GeV}.
\end{equation}
 
\subsubsection{Adiabaticity}  

Observations seem to indicate that the primordial perturbations were adiabatic.  There is no evidence for any of the physically plausible isocurvature modes, and the observational upper limits are at the few percent level \cite{Ade:2015xua}. Adiabaticity requires that the dark matter density $\rho_{dm}(x)$ and the radiation density $\rho_{rad}(x)$ vary in lock-step from point to point in such a way that:
\begin{equation}
  \label{adiabaticity}
  \frac{\delta\rho_{dm}({\bf x},t)}{\bar{\rho}_{dm}(t)}=\frac{3}{4}\frac{\delta\rho_{rad}({\bf x},t)}{\bar{\rho}_{rad}(t)}.
\end{equation}
To see that our mechanism achieves exactly this relation, consider Eq.~(\ref{Ydm}).  This equation says that the predicted dark matter number density $n_{dm}$ is proportional to the radiation entropy density $s$, with a proportionality constant that only depends on the effective number of relativistic degrees of freedom $g_{\ast}$, and the mass of the dark matter particle.  This equation was derived assuming $s$ was homogeneous, but if we imagine that $s$ varies slowly from point to point, then in a local region of the universe where $s$ (and hence $\rho_{rad}\propto s^{4/3}$) is slightly higher, this equation predicts that the local dark matter number density $n_{dm}$ (and hence the dark matter energy density $\rho_{dm}\propto n_{dm}$) is slightly higher as well, with precisely the desired excess $(\delta\rho_{dm})/\rho_{dm}=\frac{3}{4}(\delta\rho_{rad})/\rho_{rad}$ required for adiabaticity.

\subsubsection{Coldness}  

In our scenario, the dark matter particles were never in thermal contact with the radiation bath, but one can check that they were born non-relativistic: {\it i.e.} they are only created on comoving wavenumbers $k^2<\gamma$, and by the time such particles enter the horizon, their kinetic energy is already subdominant compared to their rest mass (and they just continue to get more and more non-relativistic after horizon re-entry).  Thus, these neutrinos are automatically an extremely cold form of dark matter. 

{\it Purely gravitational interactions.}  As discussed above, the same $\mathbb{Z}_{2}$ symmetry that stabilizes the dark-matter neutrino also forbids its couplings to the other standard model fields (so that it only interacts with gravity).  This means that, if this $\mathbb{Z}_{2}$ symmetry is exact, we neatly explain why the dark matter is only seen via its gravitational effects, and why a series of increasingly sensitive direct-detection and indirect-detection experiments have so far failed to detect any dark matter particles in other (non-gravitational) ways.  Unfortunately, this also means that it is hopeless to detect this dark matter candidate, directly or indirectly, using any currently-imagined non-gravitational detection scheme.

It is worth adding that the $\mathbb{Z}_{2}$ might be only approximate: in order for the dark matter predictions derived in this section to be valid, the non-gravitational couplings of the dark matter neutrinos do not have to strictly vanish -- they merely must be small enough that the dark matter remains essentially decoupled from the thermal bath in the early universe, and has a lifetime that is long relative to the current Hubble time.  This opens up the possibility that our dark matter neutrino might be indirectly detected via its decay.  (Indeed, this may have already happened: see the note added to the Discussion, regarding Ref.~\cite{Anchordoqui:2018ucj}.) 

\subsection{Self-consistency}

Let us add two remarks concerning the self-consistency of our dark matter calculations:
\begin{itemize}
\item From Eq.~(\ref{wkb}): for a fixed wavenumber $p$, the dimensionless WKB parameter $|\omega'/\omega^{2}|$ reaches a maximum at a conformal time $\tau_{p}=p/(2y_{m,1}\rho_{1}^{1/2})$ before and after the bang; and, at its peak, $|\omega'/\omega^{2}|$ is only $\gtrsim 1$ ({\it i.e.}\ the dark matter neutrino $N_{1}$ is only produced) if $p\lesssim y_{m}^{1/2}\rho_{1}^{1/4}$.   These facts together imply that $N_{1}$'s are produced at a characteristic conformal time $\tau_{\ast}\sim1/(y_{m}^{1/2}\rho_{1}^{1/4})$ before and after the bang.  If we re-express this in terms of the traditional "physical" FRW time coordinate $t$ ({\it i.e.} the proper time of a comoving observer in the Weyl gauge where the $N_{1}$'s mass is constant), it says that the $N_{1}$'s are produced at a characteristic time $t_{\ast}\sim 1/M_{1}$ before and after the Bang ({\it i.e.}\ when the age of the universe is roughly equal to their "Compton period").  Our above calculation of the dark matter abundance in Subsection \ref{Neutrino_DM} is only consistent if, during this time period ($-t_{\ast}<t<t_{\ast}$), the $N_{1}$ may be treated as a free particle, interacting only with gravity, and undisturbed by other interactions and scattering events.  This condition is satisfied for right-handed neutrinos, but {\it not} for the other particle species in the standard model (which all experience their first gauge-boson interaction long before the age of the universe reaches their Compton period).
\item As we have just seen, the $N_{1}$'s are produced at a characteristic conformal time $\tau_{\ast}\sim1/(y_{m}^{1/2}\rho_{1}^{1/4})$ before and after the bang.  At this time, the ratio of $\rho_{N}$ (the $N_{1}$ energy density predicted by our mechanism) to $\rho_{rad}$ (the radiation density) is
\begin{equation}
  \frac{\rho_{N}}{\rho_{rad}}\approx (M_{1}/\hat{\mu})^{2}.
\end{equation}
In other words: since the mass $M_{1}$ of the $N_{1}$ particle is far below the Planck scale $\hat{\mu}$, these $N_{1}$'s are a very subdominant contribution to the cosmic energy budget when they are born.  This is a self-consistency check for our calculation (since we treat the neutrinos as living on a background driven by the radiation density, while neglect the neutrino back-reaction on the cosmic expansion near the bang).
\end{itemize}

\subsection{Other predictions}

Let us mention a few other predictions that follow from our scenario:

\subsubsection{Light neutrinos are majorana; one is massless}  

First, in our scenario, the three light neutrinos obtain their masses by the usual see-saw mechanism, and hence are majorana particles (a prediction that will be tested by future experiments, including searches for neutrinoless double beta decay \cite{GomezCadenas:2011it}).  

Second, since the first row of $Y_{\nu}$ vanishes (to guarantee the stability of the dark matter particle $N_{1}$), the neutrino Dirac mass matrix $M_{D}=\langle h \rangle Y_{\nu}$ has vanishing determinant, and hence the light-neutrino seesaw mass matrix $M_{D}^{T}M^{-1}M_{D}$ does, too, which implies that the one of the three light neutrinos must be massless.  

Thus, the sum of the three light neutrino masses $m_{{\rm tot}}=\sum_{i=1}^{3}m_{i}$ should be as small as possible, given the mass-difference constraints from neutrino oscillations.  In other words, the $m_{{\rm tot}}$ must be $\sim0.05~{\rm eV}$ ($\sim 0.10~{\rm eV}$) in the normal (inverted) hierarchy.  This prediction will be tested by future cosmological observations.  (The current observational upper limit is $m_{{\rm tot}}<0.23~{\rm ev}$ \cite{Ade:2015xua}.)

If the sum of the three light neutrino masses is found to be one of these two minimal values, this will be important evidence in favor of this dark matter candidate.  If, instead, the sum is found to be anything else, this will be an important milestone, as it will rule out the last remaining dark matter candidate in the standard model (including a right-handed neutrino in each generation).

\subsubsection{Thermal leptogenesis}

So far, we have focused on the stable right-handed neutrino $N_{1}$.  The other two right-handed neutrinos, $N_{2}$ and $N_{3}$, can neatly account for the observed cosmological matter/anti-matter asymmetry via leptogenesis: in other words, due to $CP$ violation in the neutrino sector, the $N_{2}$ and $N_{3}$ particles can decay to give a lepton asymmetry, which is then converted to a baryon asymmetry by sphaleron processes above the electroweak phase transition.

These neutrinos will also be created by the mismatch of the CPT-invariant vacuum and the late time vacuum and their abundances can be calculated in a similar way to $N_1$. Explicitly we have
\begin{equation}
  \label{Y23}
  Y_{2,3}=\frac{3I}{2\pi^{2}}\left(\frac{15}{g_{\ast}}\right)^{1/4}\left(\frac{M_{2,3}}{\hat{\mu}}\right)^{3/2}
\end{equation} 
for the primordial yield.

However, at temperatures above their mass, these neutrinos have unsuppressed interactions with the thermal bath and quickly equilibrate with it, washing out any evidence of this primordial abundance.  We have run numerical simulations and confirmed that this washout is effective for all masses ${M_{2,3}\lesssim O(M_P)}$. By the time the temperature drops below the neutrino's mass and the interactions freeze out the abundance is identical to the standard thermal scenario and no evidence of the primordial abundance remains. Thus, we expect the usual thermal leptogenesis predictions to hold: see Section 4.2 in Ref. \cite{Buchmuller:2005eh}.

\subsubsection{No primordial long-wavelength gravitational waves} 

Whether a particle experiences a non-trivial Bogoliubov transformation across the bang depends, not on its spin, but on whether it is massive or massless (see Appendix \ref{exact_bogoliubov}).  Since gravitational waves are massless spin-two fields, the Bogoliubov transformation relating the pre-bang modes to the post-bang modes is trivial.  In this case, the "in" vacuum, the "out" vacuum and the $CPT$-invariant vacuum are all the same.  Thus, we predict that there are no primordial long-wavelength gravitational waves (and explain why no such waves have been detected thus far, by increasingly sensitive searches).

\section{Discussion}

We begin with a brief summary of our results.  In Sections 2 and 3, we showed how to construct the preferred $CPT$ invariant vacuum on an FRW background with time-reversal symmetry.  (We carried out the construction for spin 0 and spin 1/2, but the extension to arbitrary spin is straightforward.)  Then, in Section 4, we explained how this construction applies to our own universe -- {\it i.e.} to the standard model of particle physics (with a right-handed neutrino in each generation), living on an FRW background that is radiation-dominated near the bang.  The only dark matter candidate in this model is the right-handed neutrino $\nu_{R,1}$: if this particle is stable, it implies the Lagrangian has a $\mathbb{Z}_{2}$ symmetry under $\nu_{R,1}\to-\nu_{R,1}$ which eliminates all of the $\nu_{R,1}$'s non-gravitational couplings (so that it is completely decoupled from the thermal bath).  Nevertheless, if we assume the universe is in its $CPT$-invariant vacuum, it follows that this completely decoupled neutrino has a non-zero abundance (according to late-time observers like us); and, in fact, it neatly accounts for the observed dark matter if its mass is $4.8\times10^{8}~{\rm GeV}$.  We point out several other predictions that follow from this scenario: the light neutrinos must be Majorana, the lightest one must be massless, the matter/anti-matter asymmetry is accounted for by thermal leptogenesis, and there are no primordial long-wavelength gravitational waves.

Let us end with several remarks:
\begin{itemize}
\item There is an intriguing relationship between the stability of the dark matter neutrino, the lightness of the up quark, and the strong CP problem: see \cite{Boyle:2018tzc}.  
\item The sign flip in the fermion mass as we cross the bang (in the time direction) is an interesting temporal analogue of the sign flip in the fermion mass as we cross (in the spatial direction) the boundary separating two distinct topological phases (see \cite{Kaplan:1992bt, Kitaev:2009mg}).  In the latter case, one finds gapless modes living on the boundary \cite{Kaplan:1992bt, Rubakov:1983bb}.  It is interesting to consider what the analogous statement is for the Big Bang surface.
\item Our picture, where the regions before and after the bang are related by $CPT$, is also an interesting temporal analogue of the eternal ADS black hole (and its thermofield double state \cite{Maldacena:2001kr}), where the black hole's two exterior regions are related by $CPT$.
\item An important question is whether current observations allow the standard model to remain valid up to the Planck energy scale: see \cite{Boyle:2018tzc} for more discussion.  Even if future observations make this untenable, and force us to add new fields below the Planck scale, the basic idea introduced here of following the cosmological solution through the bang, imposing CPT, and then noticing that we can thereby explain certain features of our universe and predict a non-zero cosmological abundance of a stable massive particle, even if it is completely decoupled from all other particles, remains valid.
\item Let us mention a different perspective on our picture: in order for the two halves of spacetime (before and after the bang) to be related by $CPT$, the spatial vierbeins $e_{i}^{j}$, and hence their determinant $a^{3}={\rm det}[e_{i}^{j}]$, must flip sign as we cross the bang.  Hence, by continuity, the scale $a(\tau)$ must pass through zero in the middle (or else make an excursion into the complex plane).  If this picture is correct, then attempts to desingularize the bang by making $a(\tau)$ bounce at a finite (non-zero) minimum radius are misguided.  In our picture, the bang is a special surface in spacetime -- the surface fixed by $CPT$; and it is desingularized in a different way, via Weyl transformation (see \cite{Bars:2011mh, Bars:2011aa, Bars:2012mt, Bars:2013yba, Bars:2013vba, Gielen:2015uaa, Gielen:2016fdb} for related earlier ideas about passing through the bang).  It is interesting to consider the possibility that the low entropy (and, in particular, the low gravitational entropy) of the early universe may be explained by the requirement that the $CPT$-invariant surface is non-singular in this sense \cite{Boyle:2018tzc}.  This is reminiscent of an old suggestion by Penrose \cite{Penrose}.

\item In this initial paper, we do not yet attempt to explain certain observed features of the universe that are sometimes attributed to inflation, including spatial homogeneity, isotropy, and flatness, and the primordial scalar power spectrum.  However, our approach {\it does} already explain several {\it other} observed facts about the primordial scalar, vector and tensor perturbations \cite{Boyle:2018tzc}.  Moreover, since our description of cosmic history includes a pre-bang phase (the $CPT$ image of the post-bang phase), there is no horizon problem, nor any causality constraint preventing one from generating a scale invariant spectrum of fluctuations around a flat FRW background.  

In fact, one could even imagine adding an early inflationary phase to our story (a de Sitter like neck which connects the contracting phase to the expanding phase, while preserving the time-reversal isometry), to render the universe homogeneous, isotropic and spatially flat, and to generate the primordial density perturbations by the usual inflationary mechanism.  

If the Hubble scale $H_{I}$ during inflation is higher than the mass $M_{1}$ of the stable heavy neutrino, a non-trivial Bogoliubov transformation will again be generated between the $CPT$-symmetric and late-time vacua (due to the neutrino modes exiting and re-entering the horizon), so that a cosmological abundance of this particle will again be generated.  To estimate the abundance of such particles, we can use essentially the same argument as in the "Neutrino Dark Matter" section of Ref.~\cite{Boyle:2018tzc}.  Just as in that case, the Bogoliubov transformation is trivial ($|\beta(k)|\sim 0$) for $k>k_{cut}$ and maximal ($|\beta(k)|\sim1$) for $k<k_{cut}$; except in this case, instead of $k_{cut}=\gamma^{1/2}$, we have $k_{cut}=(M_{1}/H_{I})^{2/3}k_{{\rm end}}$, where $k_{{\rm end}}/a_{{\rm end}}=H_{I}$ is the Hubble wavenumber at the end of inflation.  Via this line of reasoning, we find that, in order to match the observed dark matter density today, the mass $M_{1}$ of the dark matter neutrino must be:
\begin{equation}
  \label{inflationary_M1}
  M_{1}\!\sim\!(2\pi)\!\!\left[\!\frac{\rho_{dm,0}}{s_{0}}m_{pl}^{3/2}H_{I}^{1/2}\!\left(\frac{\rho_{I}}{\rho_{re}}\right)^{\!\!\frac{1-3w}{4(1+w)}}\right]^{\!1/3}\!\!\!\!\!\!,\!\!
\end{equation}
where $s_{0}\sim 2.3\times10^{-38}{\rm GeV}^{3}$ is the present entropy density (in the CMB and neutrinos)  \cite{KolbTurner}, $\rho_{dm,0}\sim 9.7\times10^{-48}{\rm GeV}^{4}$ is the present dark matter energy density \cite{Ade:2015xua}, $m_{pl}=(8\pi G/3)^{-1/2}\sim 4\times10^{18}{\rm GeV}$, $\rho_{I}$ is the energy density during inflation (or, more correctly, at the end of inflation), $\rho_{re}$ is the energy density at the beginning of the radiation era, and $w$ is the effective equation of state during the reheating epoch (between the end of inflation and the start of the radiation era).    If this is the mechanism by which the dark matter is produced, $H_{I}$ must lie in the range $M_{I}\lesssim H_{I}\lesssim H_{I}^{max}$, where the upper bound $H_{I}^{max}\sim 10^{-5}m_{pl}$ comes from the non-detection of primordial gravitational waves in searches for B-mode polarization of the CMB \cite{Akrami:2018odb}.  If we push $H_{I}$ to the lower end of its allowed range, and assume instantaneous reheating (corresponding to $\rho_{I}=\rho_{re}$ or $w=1/3$), we recover our earlier prediction (\ref{M_1}) for $M_{1}$: $M_{1}=4.8\times10^{8}~{\rm GeV}$.  If we push $H_{I}$ to the upper end of its allowed range (again assuming instantaneous reheating) this pushes the predicted value of $M_{1}$ up by an order of magnitude.  And, regardless of the value of $H_{I}$, the effect of non-instantaneous reheating ({\it i.e.}\ $w<1/3$ and $\rho_{re}\ll\rho_{I}$) is to push the predicted value of $M_{1}$ a bit higher although, as may be seen from Eq.~(\ref{inflationary_M1}), the dependence on the ratio $\rho_{I}/\rho_{re}$ is rather weak.  Thus, we find that the predicted value of the neutrino dark matter mass $M_{1}$, if we add an inflationary epoch, is in the range $4.8\times10^{8}~{\rm GeV}< M_{1}<H_{I}^{max}$.

However, adding an early inflationary phase would result, as usual, in a number of additional assumptions and free parameters which we would prefer to avoid. Given the extreme economy of our explanation of the cosmic dark matter, at this point we are more interested in investigating deeper and more economical explanations of why the universe is homogeneous, isotropic and spatially flat, and how the primordial density perturbations were generated.  Follow-up work will present a new non-inflationary approach to these issues.

\item The discussion in this paper has also been restricted to the level of QFT in curved spacetime.  
Follow-up work will present a deeper viewpoint on the story presented here, inspired by the Hartle-Hawking wavefunction of the universe proposal \cite{Hartle:1983ai}.\footnote{It has been argued that the Hartle-Hawking wavefunction is $CPT$ symmetric in a certain sense \cite{Hawking:1985af}, and that it predicts that the thermodynamic arrow of time reverses at a cosmological bounce \cite{Hawking:1985af, Hartle:2011rb}.}

\item After our first paper \cite{Boyle:2018tzc} appeared on the arXiv, a follow-up paper \cite{Anchordoqui:2018ucj} suggested that the anomalous upgoing events observed by the ANITA experiment might in fact be interpreted as evidence for our dark matter candidate (the idea being that if our dark matter candidate collected in the earth over cosmic time, and then decayed with an appropriate lifetime, its decays might explain the anomalous ANITA events).  However, a subsequent analysis by the IceCube experiment \cite{IceCube:2020gbx} seems to cast doubt on this intriguing possibility.

\end{itemize}

\acknowledgments

We thank Claudio Bunster, Job Feldbrugge, Angelika Fertig, Steffen Gielen, Jaume Gomis, David B.~Kaplan, Ue-Li Pen, Laura Sberna,  Edward Witten and other participants in the workshop "The Path Integral for Gravity" at Perimeter Institute in November 2017, for valuable discussions.  Research at Perimeter Institute is supported by the Government of Canada through Innovation, Science and Economic Development, Canada and by the Province of Ontario through the Ministry of Research, Innovation and Science.

{\it Note added.}  Long after our paper appeared, it was pointed out to us that the idea that the universe before the bang could be the CPT image of the universe after the bang seems to have been first mentioned in a pair of remarkable short papers by Sakharov \cite{Sakharov1967, Sakharov1980}.

\appendix

\section{Bogoliubov transformation between ingoing and outgoing modes}
\label{exact_bogoliubov}

In this Appendix, we calculate the Bogoliubov transformation relating the ingoing modes (before the bang) to the outgoing modes (after the bang): first for a massive scalar field, and then for a massive spinor field.

In particular, we derive the result quoted in Eq.~(\ref{beta_in_out}): that the Bogoliubov $\beta$ coefficient between the ingoing and outgoing modes satisfies $|\beta({\bf p})|={\rm exp}(-\frac{\pi p^{2}}{2\gamma})$.  

We take the background FRW spacetime in the vicinity of the bang to be radiation-dominated: $a(\tau)=a_{0}\tau$ (where $a_{0}>0$ is a constant).

\subsection{Scalar field}

First consider the scalar case: we want to solve the Klein-Gordon equation (\ref{KG_eq_fourier}).

Since $a(\tau)\propto\tau$, the effective mass $\mu(\tau)=a(\tau)m$ appearing in (\ref{KG_eq_fourier}) is given by
\begin{equation}
  \mu=\gamma\tau,
\end{equation}
where $\gamma>0$ is a constant.

First consider the boundary condition (\ref{scalar_bc}) for the scalar mode functions $u_{\pm}$.  As $\tau\to\pm\infty$, we have $\omega(\tau)=[\mu^{2}+p^{2}]^{1/2}\to\pm\gamma\tau[1+\frac{1}{2}(\frac{p}{\gamma\tau})^{2}]$, so that (\ref{scalar_bc}) becomes
\begin{equation}
  \label{scalar_bc_1}
  u_{\pm}(p,\tau)\!\to\!\frac{1}{\sqrt{\pm2\gamma\tau}}{\rm exp}\!\left[\mp i \left\{\!\frac{\gamma}{2}(\tau^{2}\!-\!\tau_{0}^{2})\!+\!
  \frac{p^{2}}{2\gamma}{\rm ln}\!\left(\!\frac{\pm\tau}{\tau_{0}}\!\right)\!\right\}\right]\!.
\end{equation}

Now let us solve the Klein-Gordon equation (\ref{KG_eq_fourier}) for $u_{\pm}$.  If we define the dimensionless quantities:
\begin{equation} 
  s\equiv(2\gamma)^{1/2}\tau\qquad{\rm and}\qquad
  b\equiv\frac{p^{2}}{2\gamma},
\end{equation}
Eq.~(\ref{KG_eq_fourier}) takes the form
\begin{equation}
  \label{Phi_1}
  \frac{d^{2}\varphi}{ds^{2}}+(\frac{1}{4}s^{2}+b)\varphi=0.
\end{equation}
We can express the general solution of this equation in terms of the parabolic cylinder function $D_{p}(z)$.  $D_{p}(z)$ (where $z$ and $p$ may both be complex) is defined in Section 9.24-9.25 of Gradshteyn and Ryzhik \cite{GradshteynRyzhik}.  $D_{p}(z)$ and $D_{p}(-z)$ are two independent solutions of the differential equation (see 9.255 in \cite{GradshteynRyzhik})
\begin{equation}  
  D_{p}''(z)+(p+\frac{1}{2}-\frac{1}{4}z^{2})D_{p}(z)=0.
\end{equation}
If we define the new function 
\begin{equation}
  \label{def_f}
  f_{b}^{}(s)\equiv D_{-\frac{1}{2}+ib}(s\,{\rm e}^{-i\pi/4})
\end{equation}
we see that $f_{b}^{}(s)$, $f_{b}^{}(-s)$, $f_{b}^{\ast}(s)$ and $f_{b}^{\ast}(-s)$ are all solutions of Eq.~(\ref{Phi_1}) for $s$ real.

To see how these solutions behave in the far past or far future, we use the asymptotic expansions of $D_{p}(z)$.  In particular:
\begin{itemize}
\item for $s\to+\infty$ (${\rm arg}\,s=0$), we use formula 1 in Section 9.246 of \cite{GradshteynRyzhik} to find the asymptotic expression
\begin{subequations}
\begin{equation}
  \label{s_positive}
  f_{b}^{}(s)\approx\frac{{\rm exp}(\frac{\pi b}{4})}{s^{1/2}}{\rm exp}\left[+i\left(\frac{s^{2}}{4}+b\,{\rm ln}\,s+\frac{\pi}{8}\right)\right];
\end{equation}
\item and for $s\to-\infty$ (${\rm arg}\,s=\pi$), we use formula 2 in Section 9.246 of \cite{GradshteynRyzhik} to find the asymptotic expression
\begin{eqnarray}
  \label{s_negative}
  f_{b}^{}(s)\!\!&\!\approx\!&\!\!\frac{\,{\rm exp}(-\frac{\pi b}{4}\,)}{|s|^{1/2}}{\rm exp}\!\left[-i\!\left(\!\frac{s^{2}}{4}\!+\!b\,{\rm ln}|s|\!+\!\frac{3\pi}{8}\!\right)\!\right]\!
  \frac{i\sqrt{2\pi}}{\Gamma(\frac{1}{2}\!-\!ib)} \nonumber\\
  \!\!&\!+\!&\!\!\frac{{\rm exp}(-\frac{3\pi b}{4})}{|s|^{1/2}}{\rm exp}
  \!\left[+i\!\left(\!\frac{s^{2}}{4}\!+\!b\,{\rm ln}|s|\!-\!\frac{3\pi}{8}\!\right)\!\right].
\end{eqnarray}
\end{subequations}
\end{itemize}
From these expansions, we see that
\begin{itemize}
\item in the far future ({\it i.e.}\ for $s\to+\infty$), $f_{b}^{}(s)$ has negative frequency, and $f_{b}^{\ast}(s)$ has positive frequency; while
\item in the far past ({\it i.e.} for $s\to-\infty$), $f_{b}^{}(-s)$ has positive frequency, and $f_{b}^{\ast}(-s)$ has negative frequency.  
\end{itemize}
Thus we see that the ingoing positive frequency solution $u_{-}$ to the Klein-Gordon equation (\ref{KG_eq_fourier}) has the form
\begin{equation}
  \label{u_minus_scalar}
  u_{-}^{}(p,\tau)=c(p)f_{b}(-s).
\end{equation}
In the $\tau\to-\infty$ limit, we use the asymptotic expansion (\ref{s_positive}) to find that, in order for the expression (\ref{u_minus_scalar}) to satisfy the boundary condition (\ref{scalar_bc_1}), the coefficient $c(p)$ must be
\begin{equation}
  c(p)=\frac{{\rm exp}(-\frac{\pi b}{4})}{(2\gamma)^{1/4}}{\rm exp}\left[-i\left\{\frac{s_{0}^{2}}{4}+b\,{\rm ln}\,s_{0}+\frac{\pi}{8}\right\}\right],
\end{equation}
where $s_{0}(p)\equiv(2\gamma)^{1/2}\tau_{0}(p)$.

Now, the ingoing positive frequency solution $u_{-}^{}(p,\tau)$ can be expressed as a linear combination of the outgoing positive and negative frequency solutions, $u_{+}^{}(p,\tau)$ and $u_{+}^{\ast}(p,\tau)$, as in Eq.~(\ref{um_from_up}).  To extract the Bogoliubov coefficients $\alpha(p)$ and $\beta(p)$, we expand both sides of Eq.~(\ref{um_from_up}) in the $\tau\to+\infty$ limit: (i) we expand the left side by using the expression (\ref{u_minus_scalar}) for $u_{-}$, along with the asymptotic expansion (\ref{s_negative}); and (ii) we expand the right side by using the outgoing boundary condition (\ref{scalar_bc_1}).  By comparing these two expansions, we infer that the Bogoliubov coefficients $\alpha(p)$ and $\beta(p)$ in Eq.~(\ref{um_from_up}) are:
\begin{subequations}
  \begin{eqnarray}
    \alpha(p)\!\!&\!=\!&\!\!\frac{\sqrt{2\pi}\,{\rm exp}\left[-\frac{\pi p^{2}}{4\gamma}
    \!-i\!\left(\!\frac{s_{0}^{2}(p)}{2}\!+\!\frac{p^{2}}{\gamma}{\rm ln}(s_{0}(p))\!\right)\!\right]}{\Gamma(\frac{1}{2}-i\frac{p^{2}}{2\gamma})}
    \qquad\quad \\
    \beta(p)\!\!&\!=\!&\!\!-i\,{\rm exp}(-\frac{\pi p^{2}}{2\gamma}).
  \end{eqnarray}
\end{subequations}
As expected from Section (\ref{scalar_bogoliubov_section}): (i) the coefficient $\beta(p)$ is pure imaginary; and (ii) without loss of generality, we can choose $\tau_{0}(p)$ so that $\alpha(p)$ is real and positive.  

Furthermore, we can use the identity $|\Gamma(\frac{1}{2}+iy)|^{2}=\pi/{\rm cosh}(\pi y)$ for $y$ real (see Eq. 8.332, formula 2, in Ref.~\cite{GradshteynRyzhik}) to check that $\alpha$ and $\beta$ satisfy the required constraint for a bosonic Bogoliubov transformation: $|\alpha(p)|^{2}-|\beta(p)|^{2}=1$.  

\subsection{Spinor field}

Now consider the spinor case: we want to solve the Dirac equation (\ref{Dirac_eq}) or (\ref{Dirac_N}).

Since $a(\tau)\propto\tau$, the effective mass $\mu(\tau)=a(\tau)m$ appearing in (\ref{Dirac_eq}) or (\ref{Dirac_N}) is again given by
\begin{equation}
  \mu=\gamma\tau.
\end{equation}

Without loss of generality, we focus on the case where the ingoing positive frequency mode has momentum in the $\hat{z}$ direction (${\bf p}=p\hat{z}$) and positive helicity ($h=\frac{1}{2}$): $u_{-}(p\hat{z},\frac{1}{2},\tau)$.  As we see from Eq.~(\ref{um_from_up_spinor}), this ingoing mode is a linear combination of the outgoing positive frequency mode $u_{+}(p\hat{z},\frac{1}{2},\tau)$ and the outgoing negative frequency mode $v_{+}(-p\hat{z},\frac{1}{2},\tau)$.

We begin with the boundary conditions for these spinor mode functions:

First, consider the boundary conditions for $u_{\pm}(p\hat{z},\frac{1}{2},\tau)$.  In the limit $\tau\to\pm\infty$, with $\mu=\gamma\tau$, we have $\omega(\tau)=[\mu^{2}+p^{2}]^{1/2}\to\pm\gamma\tau[1+\frac{1}{2}(\frac{p}{\gamma\tau})^{2}]$; so that the boundary condition (\ref{spinor_bc}, \ref{uhat}) says that, as $\tau\to\pm\infty$, $u_{\pm}^{}(p\hat{z},1/2,\tau)$ has the asymptotic form
\begin{subequations}
\begin{equation}
  \label{upm_bc_RadDom}
  \frac{1}{\sqrt{2}}{\rm exp}\left[\mp i\left\{\frac{\gamma}{2}
  (\tau^{2}\!-\tau_{0}^{2})\!+\!\frac{p^{2}}{2\gamma}{\rm ln}\left(\frac{\pm\tau}{\tau_{0}}\right)\right\}\right]
  \left[\!\begin{array}{c} \pm 1-\frac{p}{2\gamma\tau} \\ 0 \\ +1\pm\frac{p}{2\gamma\tau} \\ 0 \end{array}\!\right].
\end{equation}
Then we use Eq.~(\ref{upm_T_spinor}) to infer the corresponding boundary condition for $v_{\pm}(-p\hat{z},\frac{1}{2},\tau)$: as $\tau\to\pm\infty$ it has the 
asymptotic form
\begin{equation}
  \label{vpm_bc_RadDom}
  \frac{i}{\sqrt{2}}{\rm exp}\left[\pm i\left\{\frac{\gamma}{2}
  (\tau^{2}\!-\tau_{0}^{2})\!+\!\frac{p^{2}}{2\gamma}{\rm ln}\left(\frac{\pm\tau}{\tau_{0}}\right)\right\}\right]
  \left[\!\begin{array}{c}-1\mp\frac{p}{2\gamma\tau} \\ 0 \\ \pm1-\frac{p}{2\gamma\tau} \\ 0 \end{array}\!\right].
\end{equation}
\end{subequations}

Now, to obtain the mode functions $u_{\pm}$ and $v_{\pm}$, we must solve the Dirac equation (\ref{Dirac_eq}).  Our first step is to act on Eq.~(\ref{Dirac_eq}) from the left with the operator $-(i\partial\!\!\!\;\!\!\!/\,+\mu)$ to obtain
\begin{equation}
  \label{KG_spinor}
  (\partial_{\tau}^{2}-\vec{\nabla}^{2}+\mu^{2}+i\mu'\gamma^{0})\psi=0.
\end{equation}
This Klein-Gordon-like equation (\ref{KG_spinor}) has a basis of solutions of the form
\begin{equation}
  \label{KG_spinor_solns}
  \psi^{\pm}={\rm e}^{i{\bf p}{\bf x}}\chi_{p,h}^{\pm}(\tau)\Xi_{\pm}(\hat{p},h),
\end{equation}
where the function $\chi^{\pm}_{p,h}(\tau)$ is a solution of the differential equation
\begin{equation}
  \label{chi_eq}
  \chi_{p,h}^{\pm}{}''+(p^{2}+\mu^{2}\pm i\mu')\chi_{p,h}^{\pm}=0,
\end{equation}
and the time-independent 4-component Dirac spinor
\begin{equation}
  \Xi_{\pm}(\hat{p},h)=\left(\begin{array}{r} \epsilon(\hat{p},h) \\ \pm\epsilon(\hat{p},h) \end{array}\right)
\end{equation}
is an eigenvector of $\gamma^{0}$ with eigenvalue $\pm1$, while the time-independent 2-component spinor $\epsilon(\hat{p},h)$ is an eigenvector of the helicity operator $\frac{1}{2}\hat{p}\cdot\vec{\sigma}$ with eigenvalue $h=\pm1/2$:
\begin{equation}
  \left(\frac{1}{2}\hat{p}\cdot\vec{\sigma}\right)\epsilon(\hat{p},h)=h\;\epsilon(\hat{p},h).
\end{equation}
The solution to the original Dirac equation (\ref{Dirac_eq}) is then the sum of the "$+$" and "$-$" Klein-Gordon-like solutions (\ref{KG_spinor_solns})  
\begin{equation}
  \label{psi_sum}
  \psi=\psi^{+}+\psi^{-}={\rm e}^{i{\bf p}{\bf x}}[\chi_{p,h}^{+}(\tau)\Xi_{+}(\hat{p},h)+\chi_{p,h}^{-}(\tau)\Xi_{-}(\hat{p},h)]
\end{equation}
where the pair of solutions $\chi_{p,h}^{+}(\tau)$ and $\chi_{p,h}^{-}(\tau)$ must be related as follows:
\begin{subequations}
  \begin{eqnarray}
    i\chi_{p,h}^{+}{}'(\tau)-\mu \chi_{p,h}^{+}(\tau)&=&-2hp\chi_{p,h}^{-}(\tau), \\
    i\chi_{p,h}^{-}{}'(\tau)+\mu \chi_{p,h}^{-}(\tau)&=&-2hp\chi_{p,h}^{+}(\tau).
  \end{eqnarray}
\end{subequations}
If we use the fact that $\mu=\gamma\tau$, and define the dimensionless quantities
\begin{equation}
  s\equiv(2\gamma)^{1/2}\tau\qquad{\rm and}\qquad b_{\pm}\equiv\left(\frac{p^{2}}{2\gamma}\pm\frac{i}{2}\right),
\end{equation}
Eq. (\ref{chi_eq}) becomes
\begin{equation}
  \label{chi_eq_2}
  \frac{d^{2}\chi_{p,h}^{\pm}}{ds^{2}}+(\frac{1}{4}s^{2}+b_{\pm})\chi_{p,h}^{\pm}=0.
\end{equation}
This is the same as Eq.~(\ref{Phi_1}) with $b\to b_{\pm}$; so if we define $f_{b}(s)$ as in Eq.~(\ref{def_f}), we see that $f_{b_{\pm}}^{}(s)$, $f_{b_{\pm}}^{}(-s)$, $f_{b_{\mp}}^{\ast}(s)$ and $f_{b_{\mp}}^{\ast}(-s)$ are all solutions of Eq.~(\ref{chi_eq_2}) for real $s$.

We again use the expansions (\ref{s_positive}, \ref{s_negative}) to find that:
\begin{itemize}
\item in the far future ({\it i.e.}\ for $s\to+\infty$), $f_{b_{\pm}}^{}(s)$ has negative frequency, and $f_{b_{\mp}}^{\ast}(s)$ has positive frequency; while
\item in the far past ({\it i.e.} for $s\to-\infty$), $f_{b_{\pm}}^{}(-s)$ has positive frequency, and $f_{b_{\mp}}^{\ast}(-s)$ has negative frequency.  
\end{itemize}
Thus, if we combine this with Eq.~(\ref{psi_sum}), we see that the ingoing positive frequency solution of the Dirac equation (\ref{Dirac_eq}) has the form ${\rm e}^{i(p\hat{z}){\bf x}}u_{-}(p\hat{z},\frac{1}{2},\tau)$, where
\begin{equation}
  \label{u_minus_spinor}
  u_{-}(p\hat{z},\frac{1}{2},\tau)\!=\!
  c_{+}(p)f_{b_{+}}^{}\!(-s)\!\left[\!\!\begin{array}{r} 1 \\ 0 \\ +1 \\ 0 \end{array}\!\!\right]\!+
  c_{-}(p)f_{b_{-}}^{}\!(-s)\!\left[\!\!\begin{array}{r} 1 \\ 0 \\ -1 \\ 0 \end{array}\!\!\right]\!.
\end{equation}
In the $\tau\to-\infty$ limit, we use the asymptotic expansion (\ref{s_positive}) to find that, in order for the expression (\ref{u_minus_spinor}) to satisfy the boundary condition (\ref{upm_bc_RadDom}), the coefficients $c_{\pm}(p)$ must be given by
\begin{equation}
    \frac{\pm 1}{\sqrt{2}}\!\left(\!\frac{p}{(2\gamma)^{1/2}}\!\right)^{\!\!\!\!\frac{1\pm1}{2}}
    \!\!\!\!{\rm exp}\!\left[-\frac{\pi p^{2}}{8\gamma}
    \!-i\!\left\{\!\frac{s_{0}^{2}}{4}\!+\!\frac{p^{2}}{2\gamma}{\rm ln}\,s_{0}\!+\!\frac{\pi}{8}\!\pm\!\frac{\pi}{8}\!\right\}\right]
\end{equation}
where, as before, $s_{0}(p)\equiv(2\gamma)^{1/2}\tau_{0}(p)$.

Now, the ingoing positive frequency solution $u_{-}(p\hat{z},\frac{1}{2},\tau)$ can be expressed as a linear combination of the outgoing positive frequency solution $u_{+}(p\hat{z},\frac{1}{2},\tau)$ and the outgoing negative frequency solution $v_{+}(-p\hat{z},\frac{1}{2},\tau)$ as in Eq.~(\ref{um_from_up_spinor}).  To extract the Bogoliubov coefficients $\alpha$ and $\beta$, we expand both sides of Eq.~(\ref{um_from_up_spinor}) in the $\tau\to+\infty$ limit: (i) we expand the left side using the expression (\ref{u_minus_spinor}) for $u_{-}$, along with the asymptotic expansion (\ref{s_negative}); and (ii) we expand the right side using the outgoing boundary conditions (\ref{upm_bc_RadDom}, \ref{vpm_bc_RadDom}). By comparing these two expansions, we infer that the Bogoliubov coefficients $\alpha$ and $\beta$ are
\begin{widetext}
\begin{subequations}
  \begin{eqnarray}
    \alpha(p\hat{z},1/2)&=&\frac{(4\pi\gamma)^{1/2}}{p}\frac{{\rm exp}(-\frac{\pi p^{2}}{4\gamma})}{\Gamma(-i\frac{p^{2}}{2\gamma})}{\rm exp}\left[-i\left(\frac{s_{0}^{2}(p)}{2}+\frac{p^{2}}{\gamma}{\rm ln}(s_{0}(p))-\frac{\pi}{4}\right)\right] \\
    \label{beta_appendix_spinor}
    \beta(p\hat{z},1/2)&=&-i\,{\rm exp}(-\frac{\pi p^{2}}{2\gamma}).
  \end{eqnarray}
\end{subequations}
\end{widetext}
Note that $\beta$ is given by the same expression as in the scalar case, but $\alpha$ is different.

As expected from Section (\ref{spinor_bogoliubov_section}): (i) the coefficient $\beta$ is pure imaginary; and (ii) without loss of generality we can choose $\tau_{0}(p)$ so that $\alpha(p)$ is real and positive. 

Furthermore, we can use the identity $|\Gamma(iy)|^{2}=\pi/[y{\rm sinh}(\pi y)]$ (see Eq. 8.332, formula 1, in Ref.~\cite{GradshteynRyzhik}) to check that $\alpha$ and $\beta$ satisfy the required constraint for a fermionic Bogoliubov transformation: $|\alpha|^{2}+|\beta|^{2}=1$.  

Eq.~(\ref{beta_appendix_spinor}) confirms he result quoted in Eq.~(\ref{beta_in_out}): $|\beta({\bf p},h)|={\rm exp}(-\frac{\pi p^{2}}{2\gamma})$.

\end{document}